\documentclass[aps,prb,reprint,amsmath,amssymb,superscriptaddress,longbibliography,floatfix,raggedbottom]{revtex4-2}

\usepackage{graphicx}%
\usepackage{booktabs}%
\usepackage{xcolor}%

\begin{document}

\title{IDMate: Finite-temperature error bounds for window-resolved
self-consistent-field screening}

\author{Peng Kang}
\email{pengkang@buaa.edu.cn}
\affiliation{School of Materials Science and Engineering, Beihang University, Beijing 100191, China}
\affiliation{National Key Laboratory of Artificial Intelligence for Material Science, Beihang University, Beijing 100191, China}
\affiliation{Tianmushan Laboratory, Beihang University, Hangzhou 311115, China}
\author{Da Wan}
\affiliation{School of Materials Science and Engineering, Beihang University, Beijing 100191, China}
\affiliation{National Key Laboratory of Artificial Intelligence for Material Science, Beihang University, Beijing 100191, China}
\affiliation{Tianmushan Laboratory, Beihang University, Hangzhou 311115, China}
\author{Shulin Bai}
\affiliation{School of Materials Science and Engineering, Beihang University, Beijing 100191, China}
\affiliation{National Key Laboratory of Artificial Intelligence for Material Science, Beihang University, Beijing 100191, China}
\affiliation{Tianmushan Laboratory, Beihang University, Hangzhou 311115, China}
\affiliation{Center for Bioinspired Science and Technology, Hangzhou International Innovation Institute, Beihang University, Hangzhou 311115, China}
\author{Zhen Li}
\affiliation{School of Materials Science and Engineering, Beihang University, Beijing 100191, China}
\affiliation{National Key Laboratory of Artificial Intelligence for Material Science, Beihang University, Beijing 100191, China}
\affiliation{Tianmushan Laboratory, Beihang University, Hangzhou 311115, China}
\author{Yu Liu}
\affiliation{School of Materials Science and Engineering, Beihang University, Beijing 100191, China}
\affiliation{National Key Laboratory of Artificial Intelligence for Material Science, Beihang University, Beijing 100191, China}
\affiliation{Tianmushan Laboratory, Beihang University, Hangzhou 311115, China}
\author{Lei Zheng}
\email{zhenglei@buaa.edu.cn}
\affiliation{School of Materials Science and Engineering, Beihang University, Beijing 100191, China}
\affiliation{National Key Laboratory of Artificial Intelligence for Material Science, Beihang University, Beijing 100191, China}
\affiliation{Tianmushan Laboratory, Beihang University, Hangzhou 311115, China}
\author{Li-Dong Zhao}
\email{zhaolidong@buaa.edu.cn}
\affiliation{School of Materials Science and Engineering, Beihang University, Beijing 100191, China}
\affiliation{National Key Laboratory of Artificial Intelligence for Material Science, Beihang University, Beijing 100191, China}
\affiliation{Tianmushan Laboratory, Beihang University, Hangzhou 311115, China}
\affiliation{Center for Bioinspired Science and Technology, Hangzhou International Innovation Institute, Beihang University, Hangzhou 311115, China}
\author{Huibin Xu}
\affiliation{School of Materials Science and Engineering, Beihang University, Beijing 100191, China}
\affiliation{National Key Laboratory of Artificial Intelligence for Material Science, Beihang University, Beijing 100191, China}
\affiliation{Tianmushan Laboratory, Beihang University, Hangzhou 311115, China}

\begin{abstract}
We formulate a finite-temperature residual test in IDMate that bounds
window-resolved electronic errors without a spectral-gap assumption.
For a fixed Hamiltonian and exact electron number, strong convexity of the
matrix Fermi entropy bounds the density-matrix distance and free-energy error
within a selected window. The bound remains finite at spectral crossings and
extends to weighted k points with a shared chemical potential.
We also derive a distance correction for particle-number mismatch.
Across $3{,}586$ stress trials in $70$ seeded perturbation ladders,
$1{,}942$ proposals satisfy the screen with no observed violation of the
$0.05$ window-distance criterion plus its numerical allowance.
This criterion differs from the uncorrected exact-trace bound, which six
of ten historical in-loop candidates exceed at the numerical-error scale.
An accept-or-recover loop replaces ten reference-map evaluations while
meeting terminal comparison criteria in three configurations that include
oracle-subspace controls. Additional candidates built only from
preceding-iteration orbitals yield two acceptances and one abstention.
The silicon candidate has a window distance of $1.891\times10^{-13}$ but a
normalized real-space density error of $3.754\%$.
Analytic examples separate errors from complement occupations and interblock
coupling. Window-level accuracy therefore does not imply full-state accuracy;
the screen tests compressed proposal quality, independently of nonlinear
SCF convergence or net acceleration.
\end{abstract}

\maketitle

\section{Introduction}\label{sec:introduction}

Approximate electronic states can reduce the work of a Kohn--Sham
self-consistent-field (SCF) calculation \cite{hohenberg1964,kohn1965} only when they replace expensive
operations without compromising the quantities of interest.
Such states may come from a reduced basis, an incomplete eigensolution,
previous iterations, or a learned model
\cite{davidson1975,maday2008,polack2020,brockherde2017,hazra2024,song2024neuralscf}.
Their quality is often assessed through energy changes, residuals, or
agreement with a converged reference. A useful complementary question is
whether a supplied state satisfies a computable error bound before it enters
the next iteration.

This question connects electronic-structure methods with a posteriori error
estimation. Residual estimators have been developed for eigenstates, energies,
and discretized self-consistent problems
\cite{dusson2017posteriori,herbst2020posteriori,bordignon2025guaranteed}.
Practical plane-wave bounds estimate density matrices, energies, and forces
using residuals, inverse-Jacobian estimates, and a small Schur-complement
problem \cite{cances2022bounds}.
Convergence analyses of SCF and direct minimization instead address
iterations on fixed-rank projectors and their spectral-gap dependence
\cite{cances2021convergence}.
Finite-temperature Fermi-operator expansions also admit explicit error control
\cite{rubensson2012}. These methods differ in their error objects and
assumptions on the operator, discretization, and functional.
In particular, accuracy inside a reduced subspace need not imply accuracy
of the complete electronic state.

Here we study a fixed-Hamiltonian, fixed-electron-number matrix problem at
nonzero temperature \cite{mermin1965}. The matrix Fermi entropy is strongly
convex, so standard residual estimates \cite{boyd2004convex} give a
density-matrix distance bound and a free-energy bound.
We use these estimates to screen candidates in a selected spectral window,
including complex Bloch states with one chemical potential across k points.
This application avoids a gap denominator and remains well defined
at a spectral crossing for any fixed positive temperature.

The contribution is a window-resolved screening procedure, its
implementation in IDMate, and a numerical characterization of its scope.
We test an accept-or-recover SCF loop and distinguish compressed-state,
full-state, and terminal SCF errors.
The candidates are deterministic; target-converged subspaces are included
as reference-informed controls, not as an online source of unknown solutions.
We separately test historical-subspace candidates without access to the
current full-Hamiltonian solution.

\begin{figure*}[!t]
\centering
\includegraphics[width=\linewidth]{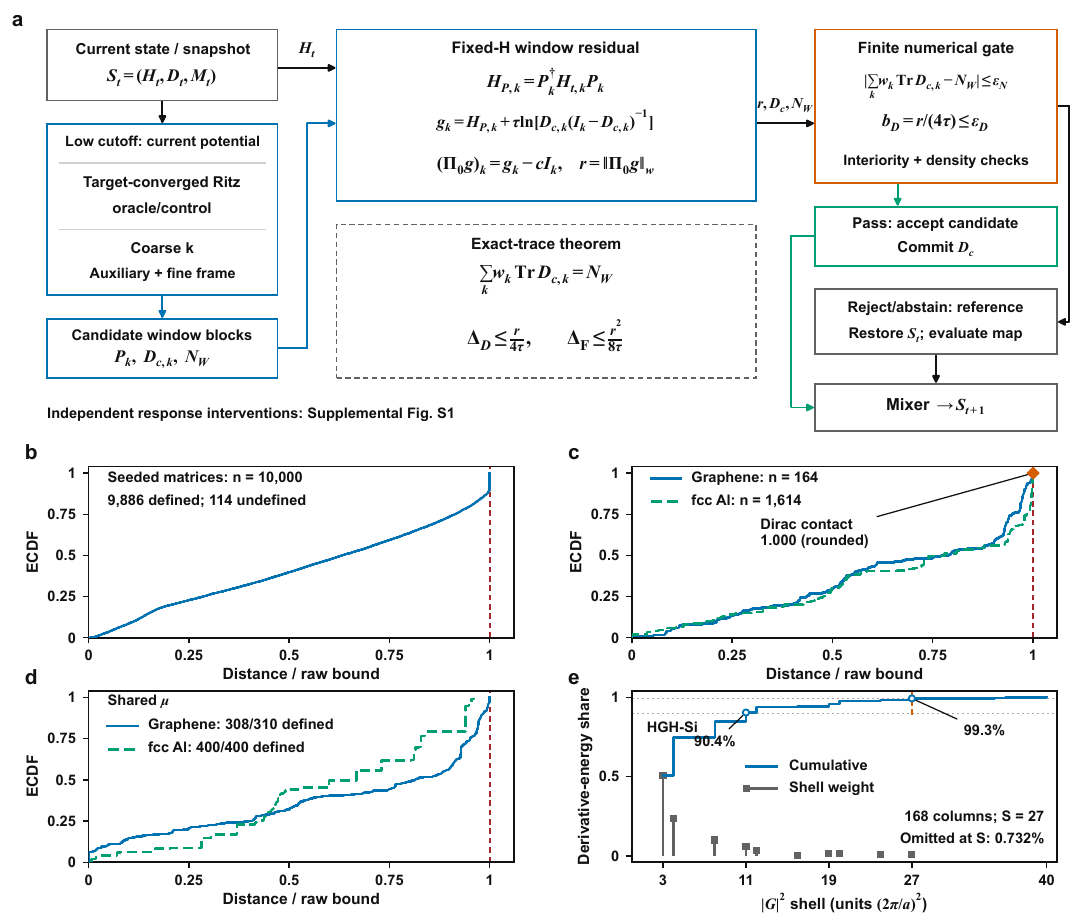}
\caption{Finite-temperature window screening and its numerical tests.
\textbf{(a)} Candidate construction, compressed residual, numerical admission
and reference-map recovery from the saved state $S_t$.
The compressed operator is $H_{P,k}=P_k^\dagger H_{t,k}P_k$; the weighted
projection uses the common constant $c$ in Eq.~\eqref{eq:weighted-projection}.
$N_W$ is the target trace after subtracting the frozen occupied complement.
$\Delta_D$ and $\Delta_F$ denote the compressed canonical errors
in Eq.~\eqref{eq:main-bounds}. The dashed theorem box assumes exact trace;
finite numerical gates are separate implementation checks.
Target-converged Ritz candidates are oracle-informed controls; coarse-k
candidates also use an auxiliary solution and a fine reference frame.
Independent response interventions appear in the Supplemental Material.
\textbf{(b)} Distance-to-bound ECDF for $9{,}886$ defined ratios from
$10{,}000$ seeded matrices of dimensions 2--8; $114$ ratios are undefined.
\textbf{(c)} Graphene Dirac-path ($n=164$) and fcc Al ($n=1{,}614$) trials.
The marked graphene maximum rounds to $1.000$, not finite-error equality.
\textbf{(d)} Shared-chemical-potential tests: $308$ defined graphene ratios
from $310$ trials, and $400$ Al ratios.
The recorded comparisons in (b)--(d) satisfy their residual-bound checks
after the stated numerical allowances. These matrix tests are separate
from both the $3{,}586$ threshold-based stress trials and the ten historical
candidates in Fig.~\ref{fig:scope}.
\textbf{(e)} An HGH-Si probe: ten retained shell weights (stems) and cumulative
weight through shell index $40$ (steps). Four shells contain $90.4\%$ and
ten shells through $S=27$ contain $99.3\%$ ($168$ columns); the omitted
weight at $S$ is $0.732\%$. This is a measured probe, not a low-rank guarantee.
AI-assisted figure preparation was checked against the numerical records
and mathematical definitions.}
\label{fig:screen}
\end{figure*}

\section{Finite-temperature window bound}\label{sec:screen}

For a Hermitian Hamiltonian $H$ of dimension $n$, temperature
$\tau=k_{\mathrm B}T>0$, and $0<N_\sigma<n$, define
\begin{equation}
\mathcal F_H(D)=\operatorname{Tr}(HD)+\tau\operatorname{Tr}
\!\left[D\ln D+(I-D)\ln(I-D)\right]
\label{eq:main-mermin}
\end{equation}
on $0\prec D\prec I$ with $\operatorname{Tr}D=N_\sigma$.
Here $D$ and $N_\sigma$ are per-spin quantities.
The unique minimizer is the canonical Fermi density matrix
$D^\star=[I+\exp((H-\mu I)/\tau)]^{-1}$, with $\mu$ fixed by the trace.
The Hamiltonian is fixed, not updated as a functional of $D$ in this minimization.

The scalar entropy curvature $[f(1-f)]^{-1}$ is at least $4$ for $0<f<1$.
Its divided-difference matrix Hessian gives the same lower bound
for noncommuting perturbations. Thus $\mathcal F_H$ is $4\tau$ strongly
convex in the Frobenius metric. Removing the constant-potential direction
gives the tangent residual
\begin{equation}
\begin{aligned}
r(D)&=\left\|\Pi_0\!\left[H+\tau\ln\!\left(D(I-D)^{-1}\right)\right]\right\|_F,\\
\Pi_0X&=X-\frac{\operatorname{Tr}X}{n}I.
\end{aligned}
\label{eq:main-residual}
\end{equation}
Standard strong-convexity estimates then imply
\begin{equation}
\begin{aligned}
\|D-D^\star\|_F&\leq\frac{r(D)}{4\tau},\\
\mathcal F_H(D)-\mathcal F_H(D^\star)&\leq\frac{r(D)^2}{8\tau}.
\end{aligned}
\label{eq:main-bounds}
\end{equation}
The proof is given in the Supplemental Material \cite{supplemental}.
These are global estimates within the stated finite-dimensional, exact-trace
problem, not for the complete nonlinear Kohn--Sham energy functional.

The constant is sharp in the small-perturbation limit near half occupation.
With $\tau=1$, $H=\operatorname{diag}(a,-a)$, and $D=I/2$, the distance
divided by its bound is $2\tanh(a/2)/a$.
It is strictly below unity for $a>0$ and tends to unity as $a\to0$.
No finite gap or nondegenerate spectrum is required.
The estimate weakens as $\tau\to0$ because its denominator vanishes.

For positive k-point weights $w_k$ summing to one, the construction extends
to $\mathcal F=\sum_k w_k\mathcal F_{H_k}(D_k)$ with one weighted trace
constraint. The corresponding norm and tangent projection are
\begin{equation}
\begin{aligned}
\|X\|_w^2&=\sum_k w_k\|X_k\|_F^2,\\
(\Pi_0X)_k&=X_k-cI_k,\qquad
c=\frac{\sum_k w_k\operatorname{Tr}X_k}{\sum_k w_k n_k}.
\end{aligned}
\label{eq:weighted-projection}
\end{equation}
For this matched functional and metric, the constant remains $4\tau$.
The minimizer uses one chemical potential across the retained blocks.

A particle-number defect can be included explicitly in the distance estimate.
For one block the additional term is $|\operatorname{Tr}D-N_\sigma|$;
for the weighted problem it is $|\delta_\sigma|/\sqrt{w_{\min}}$,
where $\delta_\sigma=\sum_k w_k\operatorname{Tr}D_k-N_\sigma$ and
$w_{\min}=\min_k w_k$.
This follows from monotonicity of canonical occupations in the chemical
potential and the triangle inequality, as proved in the Supplemental Material.
It corrects trace mismatch, not numerical errors in evaluating the residual.
For the residual-only screen, a separate electron-number check is essential:
a canonical state at the
wrong chemical potential can have zero projected residual.

An orthonormal candidate basis $P_k$ defines
$H_{P,k}=P_k^\dagger H_kP_k$. The trial block contains its per-spin
occupations; the target charge subtracts the frozen occupied complement.
The reference is the canonical state of $H_{P,k}$, generally different from
$P_k^\dagger D_k^\star P_k$ for the full Hamiltonian.
This distinction determines what the screen measures.

\section{Numerical tests}\label{sec:results}

\subsection{Matrix and material-spectrum verification}

A seeded test of $10{,}000$ real symmetric matrices found no bound violation
under its $10^{-11}$ comparison allowance. Dimensions ranged from 2 to 8,
including degenerate, near-saturated, and half-occupied cases.
The largest distance-to-bound ratios round to $1.000$, consistent with
asymptotic saturation [Fig.~\ref{fig:screen}(b)].

Material spectra test physically relevant occupation regimes.
The HGH-carbon graphene path contains a Dirac contact with residual numerical
splitting $2.19\times10^{-6}$~Ha at K.
The graphene and fcc Al spectrum-derived sets contain $164$ and $1{,}614$
trials, respectively, with no observed violations [Fig.~\ref{fig:screen}(c)].
These sets fix the electron number separately in each trial window.
The global-k tests instead impose one chemical potential across the graphene
path or the 64-point Al mesh. They contain $310$ and $400$ trials,
respectively, with no observed distance, free-energy-bound, or trace-check
violations [Fig.~\ref{fig:screen}(d)].

The proof establishes the exact-arithmetic estimate; these tests check its
numerical evaluation. Double-precision occupations can round to zero or one
at low temperature, making the spectral logarithm undefined.
The rule abstains when a retained occupation leaves the open interval.
The observed saturation floors in the temperature sweeps range from
$8.4\times10^{-3}$ to $1.38\times10^{-2}$~Ha.
These are spectrum-dependent numerical floors, not physical minimum
temperatures for DFT calculations.

\subsection{Candidate screening and reference-map fallback}\label{sec:controller}

The controller retains one mixer state and tests each candidate against
a window-distance tolerance of $0.05$.
The numerical rule also checks electron number, occupation interiority,
and the reconstructed density. Accepted candidates replace one reference-map
evaluation. Rejection or abstention uses the reference map from the
pre-proposal state. A reference-map iteration is required before declaring
SCF convergence.

Three deterministic candidates are tested. The low-cutoff candidate uses a
single eigensolution at reduced cutoff in the current effective potential,
followed by zero padding and a shared-chemical-potential occupation update.
The subspace candidate performs Rayleigh--Ritz diagonalization in orbitals
from a separately converged calculation of the target problem.
This is an oracle-assisted control: its basis is unavailable before solving
an unknown target. The coarse-k candidate transfers occupation patterns
from a separately converged coarse-mesh calculation.
It also uses a fine-mesh reference orbital frame: the current frame in
reference-scoring mode and the most recently cached frame in integrated mode.
Only the first construction uses the current potential without a converged
auxiliary or target solution.

\begin{table*}[t]
\centering
\caption{Standalone candidate tests and terminal comparisons.
Symbols denote acceptance ($\checkmark$), rejection ($\times$), and abstention
($\varnothing$). The subspace column uses target-converged orbitals.
The displayed bound is for the coarse-k candidate; its tolerance is $0.05$
in the embedded weighted metric.
The last row bypasses screening for an occupation-scrambled candidate under
a truncated iteration budget. Terminal differences concern the complete
subsequent calculation, not a consequence of the window theorem.}
\label{tab:three-zone-decisions}
\renewcommand{\arraystretch}{1.12}
\begin{tabular*}{\textwidth}{@{\extracolsep{\fill}}lccccc@{}}
\toprule
system & low cutoff & subspace (oracle) & coarse k & coarse-k bound & $|\Delta F|$ (Ha) \\
\midrule
diamond Si & $\checkmark$ & $\checkmark$ & $\times$ & $6.262$ & $8.9\times10^{-16}$ \\
graphene path & $\varnothing$ & $\checkmark$ & $\times$ & $18.224$ & $8.3\times10^{-14}$ \\
fcc Al & $\checkmark$ & $\checkmark$ & $\times$ & $4.585$ & $1.6\times10^{-14}$ \\
scrambled control & \multicolumn{3}{c}{screening bypassed} & -- & $0.132$ \\
\bottomrule
\end{tabular*}
\end{table*}

At $\tau=0.02$, $0.03$, and $0.02$~Ha for Si, graphene, and Al, respectively,
coarse-k candidates are rejected in every system
(Table~\ref{tab:three-zone-decisions}). Their bounds range from $4.585$ to
$18.224$, far above the tolerance.
The low-cutoff graphene candidate abstains because it does not supply a
usable window. The remaining accepted candidates satisfy the recorded checks.
Bypassing screening for the scrambled control gives a free-energy error
of $0.132$~Ha and a normalized density error of $0.230$.
The truncated run fails all four terminal criteria.

A larger standalone test presents $3{,}586$ trials from a gapped reference,
graphene crossing spectra, and an fcc Al spectrum.
Of these, $1{,}942$ are accepted, with no recorded violation of the
window-distance criterion plus its numerical allowance.
The remaining $1{,}644$ return to the reference branch.
Coverage among candidates whose directly measured window distances are
within tolerance is $0.814$ across the three cases.
The trials share spectra within $70$ perturbation ladders.
Treating ladders as independent, zero events give a one-sided $95\%$
upper bound of $0.042$ per ladder. This describes the tested ensemble,
not a failure probability for arbitrary materials.
The saved, count-matched replay contains finite reference distances for all
$3{,}586$ legal trials. It corroborates these counts; it does not recover
unsaved reference-solver statuses from the original aggregate-only run.

The integrated reference configurations use Si on a $2^3$ mesh at
$\tau=0.02$~Ha, graphene on a $3\times3\times1$ mesh at $0.01$~Ha,
and Al on a $3^3$ mesh at $0.02$~Ha.
These differ from the standalone configurations above.
Candidates are presented on iterations 2--7. Ten of eighteen are accepted and replace ten
reference-map evaluations. Reference-scoring mode makes the same candidate
updates but also evaluates skipped maps for comparison.
The two modes produce identical mixed-density trajectories and terminal
states for the same build and architecture.
All final states meet the free-energy, density, and frontier-band
comparison criteria against an unassisted calculation (Methods).
With oracle-assisted candidates included, this establishes decision and
recovery behavior, not deployable unknown-target acceleration.

\subsection{Compressed accuracy and full-state error}\label{sec:semantic-boundary}

The window estimate excludes the discarded complement and its coupling to
the retained subspace. A direct example makes this omission explicit.
Take $H=\operatorname{diag}(0,0,-1,1)$, $\tau=0.05$, and
$D=\operatorname{diag}(0.5,0.5,0.05,0.95)$ with trace 2.
Keeping the first two basis states gives zero window residual and error,
but a full-space distance of $1.344$ in the unembedded per-spin Frobenius norm.
The Hamiltonian has no interblock coupling, so the discrepancy comes
entirely from the complement occupations.
A second example isolates off-diagonal coupling (Supplemental Material).
Block decoupling quantifies this second contribution: at fixed particle
number and temperature, the canonical density changes by at most
$\sqrt{2}\|QHP\|_F/(4\tau)$, where $Q=I-PP^\dagger$.
Occupations in both blocks must still be controlled; the derivation and
independent matrix tests are given in the Supplemental Material.

\begin{figure*}[t]
\centering
\includegraphics[width=\linewidth]{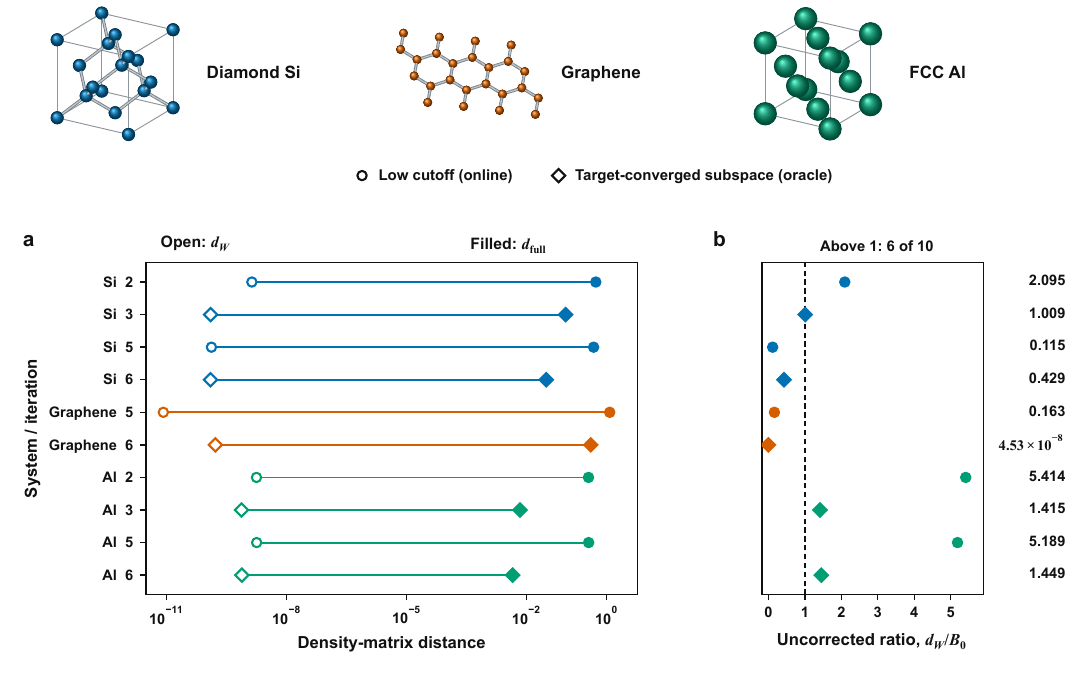}
\caption{Scope of the window criterion in ten accepted SCF candidates.
The structure motifs identify diamond Si, monolayer graphene and fcc Al;
colors match the corresponding data rows. Periodic repeats are shown for
visibility and do not indicate calculation-cell sizes. Sticks mark nearest
neighbours in Si and graphene; Al is shown as spheres.
\textbf{(a)} Paired window and full density-matrix errors for Si ($n=4$),
graphene ($n=2$), and Al ($n=4$), joined by horizontal segments within
each candidate row. Open symbols denote window errors and filled symbols
denote full-state errors. Circles identify current-potential low-cutoff
candidates; diamonds identify target-converged-subspace controls.
Each candidate type contributes five rows, labeled by system and SCF iteration.
Both distances use $[2\sum_k w_k\|\Delta D_k\|_F^2]^{1/2}$ for complex
per-spin matrices. Window errors span $8.19\times10^{-12}$ to
$1.77\times10^{-9}$, while full errors span $4.50\times10^{-3}$ to $1.202$.
\textbf{(b)} Window error divided by its uncorrected residual bound
$B_0=r_E/(4\tau)$ for the same
ten candidates; the dashed line marks unity. Six ratios exceed unity
in these inexact-trace records. The positive excesses are below the recorded
trace deviations, an empirical observation rather than a general allowance
theorem. This panel does not use the particle-number-corrected $B_W$
reported for the separate candidates in Table~\ref{tab:historical-materials}.
All ten window errors satisfy the $0.05$ acceptance threshold.
Counts refer to candidates, not independent material samples.
AI-assisted plotting was checked against the stored records.}
\label{fig:scope}
\end{figure*}

The integrated calculations show this distinction with actual candidates.
Across ten accepted steps, full density-matrix distances range from
$4.5\times10^{-3}$ to $1.202$, whereas window distances remain below
$1.8\times10^{-9}$ (Fig.~\ref{fig:scope}).
A Ritz candidate can be nearly canonical within its compressed Hamiltonian
while differing substantially from the full canonical state.
A near-zero compressed residual therefore does not alone measure subspace
quality. Six of these ten distances exceed the raw residual bound; the
largest ratio is $5.414$, with maximum positive excess
$1.429\times10^{-9}$. The Supplemental Material lists the raw bound,
distance, trace deviation, and ratio for each candidate.
Each positive excess is smaller than the trace deviation of that same row.
This numerical comparison neither establishes the cause of each excess
nor proves an allowance for arbitrary weights.
All ten distances remain below the separate acceptance threshold of $0.05$.

An enlarged set adds single-candidate insertions at different iteration
positions. Together with the initial tests, it contains 58 candidates with
defined window distances: 40 accepted and 18 rejected; eight others abstain.
The bound decision agrees with the measured window criterion on all 58.
Full-space distances, however, overlap between the accepted and rejected
groups over $[0.325,0.952]$.
The largest accepted full-space error, $1.202$, exceeds the largest rejected
value of $0.952$.
The probability that a rejected candidate has larger full-space error than
an accepted candidate is $0.694$ (AUC; $95\%$ candidate-resampling interval
$[0.544,0.831]$).
The corresponding within-material values are $0.500$, $0.250$, and $1.000$
for Si, graphene, and Al, respectively; all rejected candidates are coarse-k
and all accepted candidates are low-cutoff or subspace candidates.
This material- and candidate-dependent association does not establish a
general ranking of full-state quality. The Supplemental Material separates
same-material and cross-material pairs; the interval does not quantify
uncertainty across materials.

Terminal agreement is an empirical property of the subsequent SCF iterations,
not an extension of the fixed-$H$ bound to nonlinear convergence.
The window test, recovery procedure, and final physical accuracy can thus
be evaluated separately without assigning all three the same guarantee.

\begin{table*}[!t]
\centering
\caption{Historical-subspace candidates at the second SCF input.
Each row uses only the preceding map's orbitals, without a target-converged
subspace. $B_W$ includes the trace-defect correction; $d_W$ and $d_{\rm full}$
use the embedded weighted density-matrix norm.
$\epsilon_\rho=\sqrt{\Omega\int(\Delta\rho)^2\,d\mathbf r}/N_e$ is the
normalized real-space density error. The window threshold is $0.05$.
Graphene has no levels within $\pm18\tau$ at some k points and therefore
abstains; its complete-state comparison remains available.
Each candidate is compared with one fixed-Hamiltonian reference, not a
converged nonlinear SCF endpoint.}
\label{tab:historical-materials}
\renewcommand{\arraystretch}{1.12}
\begin{tabular*}{\textwidth}{@{\extracolsep{\fill}}llcccc@{}}
\toprule
system & decision & $B_W$ & $d_W$ & $d_{\rm full}$ & $\epsilon_\rho$ \\
\midrule
diamond Si & accept & $5.283\times10^{-10}$ & $1.891\times10^{-13}$ & $0.100$ & $0.038$ \\
graphene & abstain & -- & -- & $0.095$ & $0.060$ \\
fcc Al & accept & $1.414\times10^{-10}$ & $4.120\times10^{-12}$ & $3.489\times10^{-3}$ & $1.202\times10^{-3}$ \\
\bottomrule
\end{tabular*}
\end{table*}

\subsection{Historical-subspace material candidates}\label{sec:historical-materials}

To remove dependence on target-converged orbitals, we constructed candidates
from the immediately preceding reference map in Si, graphene, and Al.
Starting from a neutral uniform density, the first map supplied the orbitals
for a Rayleigh--Ritz calculation at the next mixed input.
The candidate and its screening decision were saved before computing the
complete reference state of that same Hamiltonian.
These tests used the particle-number-corrected distance bound and retained
the actual occupations outside the selected window (Methods).

Si and Al passed the window criterion, whereas graphene abstained because
the prescribed window was unavailable (Table~\ref{tab:historical-materials}).
The accepted Si candidate had a full-state distance of $0.100$ and a
normalized density error of $0.038$, despite negligible compressed error.
Al gave smaller complete-state and density errors, while the graphene
candidate remained measurable after abstention.
The small compressed residuals are expected from the Ritz construction;
the distinguishing quantities are the remaining full-state error and
the availability of the prescribed window.
Thus the distinction between window and full-state accuracy also occurs
without target-solution information.
Two reference maps per material completed this comparison; no subsequent
nonlinear SCF iteration was propagated.

\subsection{Computational work}\label{sec:cost}

Once the candidate and compressed Hamiltonian are available, the screen
acts on the window rather than the full plane-wave space.
The real-embedded implementation uses dense spectral operations with cubic
cost in the embedded dimension. Candidate generation, projection, and
density reconstruction can nevertheless dominate the work.
The bound's lower dimension alone does not imply a lower total cost.

In silicon at 25~Ha, 64 k points, and approximately $1.6\times10^3$ plane waves,
the integrated run converges in 193 iterations.
Four of six candidates replace reference calls; two coarse-k candidates
are rejected. The historical composite cost ratio is $7.870$
relative to its all-reference comparison.
Its accounting contains overlapping candidate and screening intervals,
and target-reference preparation is not charged as online candidate work.
The Supplemental Material retains this implementation record, not a
corrected wall-time ratio or an acceleration measurement.
An isolated optimized-screen benchmark is also distinct from the complete
candidate-and-recovery path.

Candidates were presented only on iterations 2--7, so the replaceable
fraction is bounded by this schedule rather than by the method itself.
With equal-cost reference calls, an unchanged trajectory, and free candidate
processing, replacing four of 193 calls yields only $193/189=1.021$
times acceleration. Even replacing all six scheduled calls would give
only $193/187=1.032$ under the same assumptions.
Actual gains require a larger replaceable fraction and
cheap candidates, including all preparation and recovery work.
The present experiments establish functional integration, not net speedup.

\subsection{Physical implementation checks}\label{sec:anchor}

IDMate uses a norm-conserving HGH operator for the material tests.
We assess silicon through cutoff and k-mesh convergence, a PySCF comparison
with matching pseudopotential parameters, and a separate VASP-PAW shape
comparison \cite{kresse1996}. These tests answer different questions.

All fourteen silicon cutoff-by-k-mesh calculations converge.
At 25~Ha, successive k-mesh energy changes satisfy the $1$~meV/atom
criterion, whereas the $20\to25$~Ha cutoff change remains approximately
$2.46$--$2.47$~meV/atom (Table~\ref{tab:physical-convergence}).
The selected calculation therefore retains a measurable cutoff error.

The PySCF GTH-PADE parameters match the analytic HGH silicon operator,
including its off-diagonal projector coefficient \cite{sun2020pyscf,hartwigsen1998}.
At the common $\tau=0.020$~Ha and half-shifted $4^3$ mesh, the stored gaps
are $41.640$ and $43.320$~mHa for IDMate and PySCF, respectively.
Their difference is $1.680$~mHa, or $45.711$~meV, using unrounded values.
The IDMate 10--25 Ha cutoff ladder decreases from $41.949$ to $41.640$~mHa
and does not approach this PySCF value monotonically.
The different plane-wave and Gaussian bases remain relevant;
the residual is not assigned to a single cause.

\begin{table}[tbp]
\centering
\caption{Selected silicon convergence and same-operator checks.
Energy changes are per atom, with a $1$~meV/atom convergence target.
The gap comparison uses $\tau=0.020$~Ha and half-shifted $4^3$ meshes
in both implementations, with different basis representations.
The last row is a diagnostic, not a convergence criterion.}
\label{tab:physical-convergence}
\renewcommand{\arraystretch}{1.13}
\begin{tabular*}{\columnwidth}{@{\extracolsep{\fill}}lc@{}}
\toprule
comparison & difference \\
\midrule
25 Ha, $4^3\to6^3$ mesh & $0.122$ meV/atom \\
25 Ha, $6^3\to8^3$ mesh & $8.47\times10^{-4}$ meV/atom \\
$20\to25$ Ha, $4^3$ mesh & $2.468$ meV/atom \\
IDMate--PySCF gap, $4^3$ mesh & $1.680$ mHa \\
\bottomrule
\end{tabular*}
\end{table}

\begin{figure*}[tp]
\centering
\includegraphics[width=\textwidth]{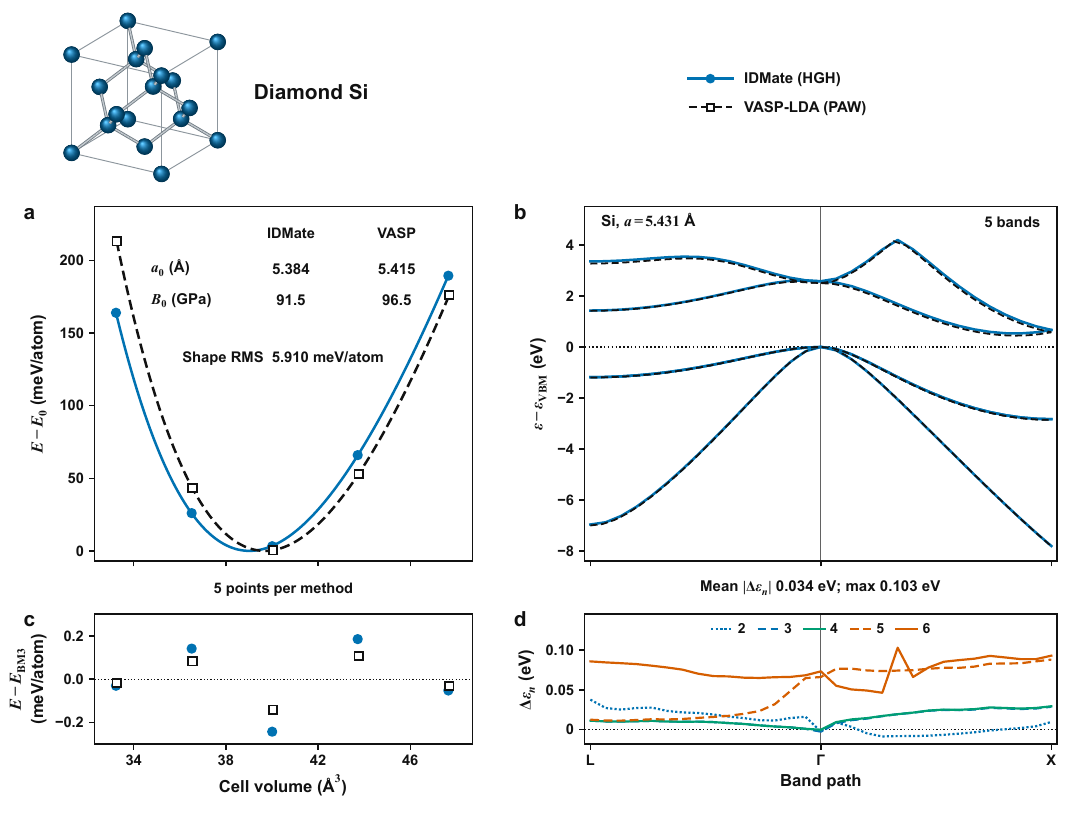}
\caption{Silicon comparisons with VASP-LDA(PAW).
The motif is the conventional diamond cell corresponding to the two-atom
primitive band-calculation cell.
\textbf{(a)} Five equation-of-state points per method and third-order
Birch--Murnaghan fits, relative to separate equilibrium energies.
IDMate gives $a_0=5.384$~\AA{}, $B_0=91.5$~GPa; VASP gives
$5.415$~\AA{}, $96.5$~GPa. The shape RMS difference is $5.910$~meV/atom
at 41 common scaled volumes $V/V_0$, not equal raw volumes.
\textbf{(b)} Five VBM-aligned bands at $a=5.431$~\AA{}, paired by band-index
offset from the VBM along $\mathrm{L}$--$\Gamma$--$\mathrm{X}$.
\textbf{(c)} Signed residuals from each method's own energy fit.
\textbf{(d)} Signed band differences
$\Delta\epsilon_n=\epsilon_n^{\mathrm{IDMate}}-\epsilon_n^{\mathrm{VASP}}$
after VBM alignment and linear interpolation of VASP values in reciprocal
distance onto the IDMate grid: 31 points per band, 155 pairs, mean absolute
difference $0.034$~eV, and maximum absolute difference $0.103$~eV.
Both band panels use segment-fractional path coordinates.
These compare shapes, not matched Hamiltonians; pseudopotential, smearing,
k-mesh, and cutoff differences are specified in Methods.
AI-assisted plotting was checked against the numerical records.}
\label{fig:physical-validity}
\end{figure*}

The HGH and PAW total-energy zeros are not directly comparable.
Figure~\ref{fig:physical-validity} therefore compares equation-of-state shapes
and VBM-aligned bands. VASP uses Gaussian smearing of $0.05$~eV and a
Gamma-centered mesh, whereas IDMate uses Fermi--Dirac occupations at
$0.020$~Ha and a half-shifted mesh.
The observed differences cannot be attributed solely to the pseudopotential.
Graphene and Al supply the crossing and metallic spectra used above;
additional reference calculations are documented in the Supplemental Material.

\section{Discussion}\label{sec:discussion}

The method supplies an a posteriori quality test for a prescribed,
finite-temperature compressed electronic problem.
Its mathematical basis is standard strong convexity; its application
specifies the electron-number constraint, k-weighted metric, complex-state
representation, and SCF recovery behavior.
Unlike a gap-denominator estimate, it remains finite at a crossing for
fixed $\tau>0$. Its scope is determined by the selected subspace,
not by the candidate generator.

This complements existing finite-temperature operator error control and
a posteriori electronic-structure estimates
\cite{rubensson2012,herbst2020posteriori,bordignon2025guaranteed}.
The window bound does not estimate the excluded-space or discretization
errors addressed by broader analyses.
Moreau--Yosida formulations regularize density functionals while retaining
an exact route to the physical energy at finite regularization parameter
\cite{kvaal2014,penz2019guaranteed,penz2020erratum}.
Here $\tau$ sets the occupation function, and the estimate applies to the
unmodified fixed-$H$ problem at that temperature.

The contrast between compressed and complete states is a substantive
outcome. Ritz candidates can satisfy the compressed canonical problem
almost exactly without accurately spanning the full reference state.
Analytic examples identify complement occupations and coupling as
independent reasons. Material calculations show the same distinction,
while subsequent reference iterations can recover the tested terminal
observables.

Established mixing methods address the nonlinear iteration, whereas the
window screen assesses a supplied state for one fixed Hamiltonian.
The Supplemental Material compares linear mixing, Kerker preconditioning,
and specified Pulay variants \cite{kerker1981,pulay1980}.
Eleven converged material--policy combinations across Si, graphene, and Al
reached the same fixed point within each material, with maximum free-energy
difference $1.4\times10^{-13}$~Ha.
These finite parameter tests establish endpoint consistency, not superiority
over optimally tuned mixing or universal fixed-point invariance.

A work-prediction study attempted $240$ trajectories using a zero-temperature
subspace diagnostic, distinct from the present residual.
The primary analysis missed its predefined $20\%$ improvement criterion;
the subsequent intercept-corrected fit gave $12.650\%$, with a
resampling interval spanning zero (Supplemental Material).
The response variable combined different kernel counters and completed-iteration
information. A later pseudopotential counter-increment analysis also failed
across held-out system categories. These results do not establish prospective
work prediction or exclude response diagnostics under a common work definition.
Separate interventions distinguish fixed-map conditioning from mixer
amplification; neither experiment derives the window bound.

Historical orbitals avoid target-solution information, but full-state errors
and window availability differ across the three materials.
Acceleration further requires reliable evaluation near occupation boundaries
and candidate costs below the work displaced. Complement-error estimates
would extend the assessment from a window to the complete electronic state.

\section{Computational details}\label{sec:methods}

\subsection{Electronic-structure calculations}

IDMate implements non-spin-polarized Kohn--Sham calculations in Rust.
Material calculations use complex Bloch states, weighted kinetic and
band sums, and density reconstruction on a periodic FFT grid.
The local-density approximation combines Slater exchange with
Perdew--Zunger correlation \cite{ceperley1980,perdew1981}.
Analytic HGH pseudopotentials use separable nonlocal projectors
\cite{goedecker1996,hartwigsen1998,kleinman1982}; ion--ion interactions
use an Ewald sum \cite{ewald1921}.
The Supplemental Material gives parameters, normalization tests, and
external pseudopotential parsing limitations.

The small integrated reference configurations use a self-adjoint dense
solver with eigenpair residual and orthonormality checks at $10^{-11}$.
The silicon cutoff comparison and production-size calculation instead use
the block Davidson path \cite{davidson1975}, with separately checked
eigenpair residuals.
Potential matrices use FFT coefficients and a cell-center phase.
Comparisons with direct quadrature are supplied separately from
independent-code physical comparisons.

The silicon PySCF calculation uses version 2.14, GTH-PADE, GTH-QZV2P,
and a 25~Ha integration-grid cutoff, not a plane-wave basis cutoff.
The reported gap comparison uses $\tau=0.020$~Ha and equivalent half-shifted
$4^3$ Monkhorst--Pack meshes \cite{monkhorst1976}.
Free energies use $F=E-\tau S$ with dimensionless electronic entropy $S$.
The same-temperature record is distinguished from the stored zero-smearing
PySCF calculation. VASP comparisons use the differing settings stated above.

\subsection{Window construction and numerical acceptance}

For each candidate k block, the algorithm counts levels within
$\pm18\tau$ of its chemical potential.
The largest count, limited by usable bands at every k point, sets a common
block size $m$. Each block then retains the $m$ bands nearest the
chemical potential. Individual retained blocks can thus extend beyond
$\pm18\tau$.
The first excluded bands must be essentially occupied below and empty above.
Retained occupations must lie inside the open interval;
frozen external states may be fully occupied or empty.

Complex blocks use the faithful real embedding
\begin{equation}
\mathcal R(A)=
\begin{pmatrix}
\operatorname{Re}A&-\operatorname{Im}A\\
\operatorname{Im}A&\operatorname{Re}A
\end{pmatrix}.
\end{equation}
The embedding obeys $\operatorname{Tr}\mathcal R(D_k)=2\operatorname{Tr}D_k$
and scales Frobenius distances by $\sqrt{2}$.
Separately, physical spin degeneracy gives $N_e=2N_\sigma$.
Thus the weighted trace of the embedded per-spin matrices equals $N_e$
numerically; no additional spin factor is applied to that trace.
The bound and measured distance use the same rescaling, leaving the
strong-convexity constant unchanged.
The threshold $0.05$ is absolute in this embedded weighted metric;
its physical stringency is not size independent.

Kernel and controller electron-number gates use $10^{-8}$ and $10^{-6}$,
respectively. Finite tolerances do not make an inexact-trace candidate
satisfy the exact-trace theorem.
Recorded comparisons also use an empirical roundoff allowance depending
on trace drift and occupation margin. It is not a proved bound for arbitrary
weights and spectra; its construction and a weighted-trace limitation
are given in the Supplemental Material.
The reconstructed density is checked for finite values, non-negativity,
and charge error below $10^{-10}\max(N_e,1)$.

For the historical-subspace tests, the retained basis ends two bands above
the last preceding-map occupation greater than $10^{-12}$.
Rayleigh--Ritz diagonalization and a common chemical potential determine
the new occupations, with zero density on the omitted orthogonal complement.
The window target subtracts the actual excluded occupations, including
fractional tails, rather than replacing them by integers.
Missing common-window bands cause abstention instead of a reduced window.
The distance bound includes $|\delta N_e|/\sqrt{2w_{\min}}$ in the embedded
metric; no empirical trace allowance is added.
Post-decision reference charges must agree within $10^{-10}$ electrons.
The fixed-$H$ energy bound is checked separately against a canonical state
at the candidate's measured trace, not against a different target trace.

A separate common-energy-offset test exposed cancellation in an earlier
uncentered floating-point residual evaluation (Supplemental Material).
The exact bound is gauge invariant, but its numerical evaluation requires
resolving the entropy gradient relative to the Hamiltonian energy scale.
Reference densities must also satisfy their electron-number constraint
before they can support an error comparison.
The corrected implementation removes an unresolved common energy origin
before adding the entropy gradient and checks the computed reference trace.

\subsection{SCF and statistical comparisons}

Reference-scoring mode evaluates the reference map even when a candidate
drives the mixer; integrated mode omits that call after acceptance.
Observation-only evaluation measures skipped-map errors without feeding
the reference state into the update.
Terminal comparisons against an unassisted converged calculation require
$|\Delta F|<10^{-7}$~Ha, normalized density distance below $10^{-5}$,
and HOMO and LUMO shifts below $10^{-5}$~Ha.
The normalized density distance is
$\sqrt{\Omega\int(\Delta\rho)^2\,d\mathbf r}/N_e$.
These concern final states, not intermediate full-space accuracy.
The historical-subspace tests instead compare single fixed-$H$ maps.
Their two next-input densities are linear combinations of the same saved
input and the respective output densities, without evaluating another map.
Reference-informed candidate generation is distinct from observation-only
measurement.
Both endpoint solves must be converged and all comparison quantities finite.
Direct synthetic tests identified and corrected missing reference-convergence
and finiteness checks in the earlier terminal decision routine
(Supplemental Material).

A window violation is defined only when a finite reference distance has
been evaluated. Missing or failed reference calculations cannot support
a zero-violation conclusion. The historical implementation and numerical
boundaries are detailed in the Supplemental Material.
Perturbations sharing a spectrum are grouped in seeded ladders.
For zero events in $n$ independent units, the exact one-sided $95\%$
binomial upper bound is $1-0.05^{1/n}$; $n=70$ gives $0.042$.
Enlarged in-loop comparisons share three system trajectories;
pooled associations are descriptive rather than estimates over materials.
The separate counter-regression analysis and conditional inclusion rules
are confined to the Supplemental Material.

OpenAI Codex assisted with implementation, numerical analysis, figure
preparation, and manuscript revision. AI-assisted changes were checked
using explicit mathematical examples, stored numerical outputs, and
source-level consistency checks.
The candidate constructions tested here are deterministic and do not
use a trained machine-learning model.

\section*{Data and code availability}
Derived numerical inputs and scripts for all main and supplemental figures,
the three-material comparison tables, and the trace, stratified-separation,
response-shell, and subspace-reuse analyses are available at
{\urlstyle{same}\url{https://github.com/kpleo/idmate_reproducibility}}
(7 September 2026 release).
The documented coverage distinguishes reconstruction of stored results
from recalculation; other tables and complete SCF replay are not fully covered.
The IDMate solver is maintained separately. Licensed pseudopotentials and
raw VASP outputs are not redistributed.

\begin{acknowledgments}
This work was supported by the National Science Fund for Distinguished Young
Scholars (No.~51925101), the National Natural Science Foundation of China
(Nos.~52450001 and 12104370), the Tianmushan Laboratory Research Project
(Nos.~TK2024D006 and TK2023C021), and the ``Pioneer'' and ``Leading Goose''
R\&D Program of Zhejiang (Project No.~2024SSYS0084). Computational resources
were provided by Tianmushan Laboratory and Beihang University.
\end{acknowledgments}

\section*{Author contributions}
P.K.: conceptualization, methodology, software, validation, formal
analysis, investigation, data curation, and writing. D.W., S.B., Z.L., and
Y.L.: validation, formal analysis, and investigation. L.Z., L.-D.Z., and H.X.:
conceptualization, supervision, funding acquisition, review, and editing.

\section*{Competing interests}
The authors declare no competing interests.

\bibliography{references}

\end{document}


\title{Supplemental Material for ``IDMate: Finite-temperature error bounds
for window-resolved self-consistent-field screening''}

\author{Peng Kang}
\email{pengkang@buaa.edu.cn}
\affiliation{School of Materials Science and Engineering, Beihang University, Beijing 100191, China}
\affiliation{National Key Laboratory of Artificial Intelligence for Material Science, Beihang University, Beijing 100191, China}
\affiliation{Tianmushan Laboratory, Beihang University, Hangzhou 311115, China}
\author{Da Wan}
\affiliation{School of Materials Science and Engineering, Beihang University, Beijing 100191, China}
\affiliation{National Key Laboratory of Artificial Intelligence for Material Science, Beihang University, Beijing 100191, China}
\affiliation{Tianmushan Laboratory, Beihang University, Hangzhou 311115, China}
\author{Shulin Bai}
\affiliation{School of Materials Science and Engineering, Beihang University, Beijing 100191, China}
\affiliation{National Key Laboratory of Artificial Intelligence for Material Science, Beihang University, Beijing 100191, China}
\affiliation{Tianmushan Laboratory, Beihang University, Hangzhou 311115, China}
\affiliation{Center for Bioinspired Science and Technology, Hangzhou International Innovation Institute, Beihang University, Hangzhou 311115, China}
\author{Zhen Li}
\affiliation{School of Materials Science and Engineering, Beihang University, Beijing 100191, China}
\affiliation{National Key Laboratory of Artificial Intelligence for Material Science, Beihang University, Beijing 100191, China}
\affiliation{Tianmushan Laboratory, Beihang University, Hangzhou 311115, China}
\author{Yu Liu}
\affiliation{School of Materials Science and Engineering, Beihang University, Beijing 100191, China}
\affiliation{National Key Laboratory of Artificial Intelligence for Material Science, Beihang University, Beijing 100191, China}
\affiliation{Tianmushan Laboratory, Beihang University, Hangzhou 311115, China}
\author{Lei Zheng}
\email{zhenglei@buaa.edu.cn}
\affiliation{School of Materials Science and Engineering, Beihang University, Beijing 100191, China}
\affiliation{National Key Laboratory of Artificial Intelligence for Material Science, Beihang University, Beijing 100191, China}
\affiliation{Tianmushan Laboratory, Beihang University, Hangzhou 311115, China}
\author{Li-Dong Zhao}
\email{zhaolidong@buaa.edu.cn}
\affiliation{School of Materials Science and Engineering, Beihang University, Beijing 100191, China}
\affiliation{National Key Laboratory of Artificial Intelligence for Material Science, Beihang University, Beijing 100191, China}
\affiliation{Tianmushan Laboratory, Beihang University, Hangzhou 311115, China}
\affiliation{Center for Bioinspired Science and Technology, Hangzhou International Innovation Institute, Beihang University, Hangzhou 311115, China}
\author{Huibin Xu}
\affiliation{School of Materials Science and Engineering, Beihang University, Beijing 100191, China}
\affiliation{National Key Laboratory of Artificial Intelligence for Material Science, Beihang University, Beijing 100191, China}
\affiliation{Tianmushan Laboratory, Beihang University, Hangzhou 311115, China}

\maketitle

\section{Mermin finite-temperature residual bound}
\label{sec:mermin}

The screening bound concerns the canonical (fixed-electron-number)
finite-temperature density matrix \cite{mermin1965}. The separate
zero-temperature projector diagnostic used in
Sec.~\ref{sec:work-prediction} is not the screening residual.
This section proves the bound for a fixed Hamiltonian; the proof does not
use the SCF Jacobian $J$, the response operator $A=I-J$, or their
rank-one interventions. The setting is finite-dimensional:
$H$ is a real symmetric
$n \times n$ Hamiltonian (any fixed discretization), $\tau > 0$ the
electronic temperature or smearing parameter, and the per-spin electron
number lies strictly between $0$ and $n$. The theorem is stated for the
per-spin density matrix $0\prec D_{\sigma}\prec I$. The spin-degenerate
implementation stores occupations $n_i\in[0,2]$ but screens the per-spin
occupations $f_i=n_i/2$, with complex blocks represented by the faithful
real embedding described below. The $4\tau$ modulus applies in this
explicit embedded per-spin metric; it is not invariant under
$P=2D_{\sigma}$ in the unscaled spatial Frobenius metric.

\paragraph{Fixed-trace Mermin problem.}
The canonical density matrix is the unique minimizer of the Mermin free
energy
\begin{equation}
\mathcal F_H(D) \;=\; \operatorname{Tr}(H D)
\;+\; \tau \, \operatorname{Tr}\!\big[ D \ln D + (I - D) \ln (I - D) \big]
\label{eq:mermin-free-energy}
\end{equation}
over symmetric $D$ with $0 \prec D \prec I$ and
$\operatorname{Tr} D = N_\sigma$.
Existence, interiority and uniqueness of the minimizer are established
after the strong-convexity proof below.

\paragraph{$4\tau$ strong convexity on the fixed-trace space.}
On the affine space $\operatorname{Tr} D = N_\sigma$, the linear term
$\operatorname{Tr}(HD)$ contributes zero curvature, so all curvature
comes from the entropy term $\operatorname{Tr} h(D)$ with
$h(x) = x \ln x + (1-x) \ln (1-x)$. We give the Hessian along an
arbitrary symmetric direction in Loewner (divided-difference) form.
Fix an admissible $D$ with spectral
decomposition $D = U \operatorname{diag}(d_1,\dots,d_n) U^{\top}$,
$d_i \in (0,1)$, and let $X$ be any real symmetric direction with
matrix elements $\tilde X_{ij} = (U^{\top} X U)_{ij}$ in the
eigenbasis of $D$. Standard second-order eigenvalue perturbation
theory applied to $D + tX$ (first-order shift $\tilde X_{ii}$ of
$d_i$; second-order shift
$\sum_{j \neq i} |\tilde X_{ij}|^{2}/(d_i - d_j)$) yields, via
$\operatorname{Tr} h(D + tX) = \sum_i h(d_i(t))$ and collecting the
ordered pair $(i,j)$, the exact Hessian
\begin{equation}
\mathrm{D}^{2}\operatorname{Tr} h(D)[X,X]
\;=\; \sum_{i,j} h^{[1]}(d_i, d_j)\, |\tilde X_{ij}|^{2},
\qquad
h^{[1]}(a,b) \;=\;
\begin{cases}
\dfrac{h'(a) - h'(b)}{a - b}, & a \neq b, \\[2.2ex]
h''(a), & a = b,
\end{cases}
\label{eq:loewner-hessian}
\end{equation}
where $h^{[1]}$ is the first divided difference of $h'$, continuous on
$(0,1) \times (0,1)$. Equation~\eqref{eq:loewner-hessian} is the
Daleckii--Krein formula \cite{daleckiikrein1965,higham2008functions}
for the Fr\'echet derivative of the matrix function $h'$ paired with the
direction $X$. No non-degeneracy
assumption is needed: at coincident eigenvalues $d_i = d_j$ the divided
difference is understood in its continuous extension
$h^{[1]}(a,a) = h''(a)$, so the Hessian formula holds for every real
symmetric $X$ without exception, including directions that do not
commute with $D$. The diagonal entries $h^{[1]}(d_i,d_i) = h''(d_i)$
are the scalar eigendirection curvature of
Eq.~\eqref{eq:strong-convexity} below. The key
estimate is a bound on $h^{[1]}$. Since $h'$ is continuously
differentiable on $(0,1)$, the fundamental theorem of calculus applied
to $t \mapsto h'(ta + (1-t)b)$ gives the integral representation
\begin{equation}
h^{[1]}(a,b)
\;=\; \frac{h'(a) - h'(b)}{a - b}
\;=\; \int_{0}^{1} h''\!\big(ta + (1-t)b\big)\,\mathrm{d}t
\qquad (a \neq b),
\label{eq:divided-difference-integral}
\end{equation}
which extends to $a = b$ by continuity (the mean value theorem form of
the same statement: $h^{[1]}(a,b) = h''(\xi)$ for some $\xi$ between
$a$ and $b$). The scalar curvature is
\begin{equation}
h''(x) \;=\; \frac{1}{x} + \frac{1}{1-x} \;=\; \frac{1}{x(1-x)}
\;\geq\; 4
\qquad \text{for all } x \in (0,1),
\label{eq:strong-convexity}
\end{equation}
because $1 - 4x(1-x) = (2x-1)^{2} \geq 0$, with equality exactly at
$x = 1/2$. Every convex combination $ta + (1-t)b$ lies in $(0,1)$, so
Eq.~\eqref{eq:divided-difference-integral} integrates a function
bounded below by $4$ and
\begin{equation}
h^{[1]}(a,b) \;\geq\; 4
\qquad \text{for all } a,b \in (0,1),
\label{eq:divided-difference-bound}
\end{equation}
with equality in the diagonal limit $a = b = 1/2$. Substituting
\eqref{eq:divided-difference-bound} into \eqref{eq:loewner-hessian} and
using the unitary invariance of the Frobenius norm,
$\sum_{i,j} |\tilde X_{ij}|^{2} = \|X\|_F^{2}$, gives
\begin{equation}
\mathrm{D}^{2}\mathcal F_H(D)[X,X]
\;=\; \tau\, \mathrm{D}^{2}\operatorname{Tr} h(D)[X,X]
\;\geq\; 4\tau \, \|X\|_F^{2}
\label{eq:strong-convexity-general}
\end{equation}
for every real symmetric $X$, in particular for every traceless $X$
tangent to the fixed-trace affine space. Hence $\mathcal F_H$ is
$4\tau$-strongly convex on the fixed-trace tangent space in the
Frobenius metric. The constant is sharp. At half occupation
$D = \tfrac12 I$ (admissible when $N_\sigma = n/2$), the integrand of
\eqref{eq:divided-difference-integral} is identically $4$ for every
direction. Equality in \eqref{eq:strong-convexity-general} is then attained
for every tangent direction. This is equality of the local Hessian bound,
not of the residual-to-distance bound at a finite nonzero error.
For example, at $\tau=1$, $H=\operatorname{diag}(a,-a)$ and $D=I/2$,
with $\operatorname{Tr}D=1$ and $a>0$,
\begin{equation}
\frac{\|D-D^\ast\|_F}{r(D)/(4\tau)}
=\frac{2\tanh(a/2)}{a}<1.
\label{eq:sharpness-limit}
\end{equation}
The ratio tends to one as $a\to0$, so the distance constant is
asymptotically sharp. The deficit from unity is $8.325\times10^{-4}$ at
$a=0.1$ and $8.333\times10^{-6}$ at $a=0.01$.
The latter ratio rounds to $1.000$ at three decimal places without giving
finite-error equality.

\paragraph{Existence, interiority and uniqueness of the minimizer.}
For $0<N_\sigma<n$, define the feasible set
\begin{equation*}
\mathcal{D}_{N_\sigma}
=\{D=D^{\top}:0\preceq D\preceq I,\ \operatorname{Tr}D=N_\sigma\}.
\end{equation*}
This set is closed, bounded, and convex in the finite-dimensional matrix space
and is therefore compact. The function $h$ extends continuously to $[0,1]$
by $h(0) = h(1) = 0$. Thus $\mathcal F_H$ is continuous on the feasible set
and attains its minimum.

No minimizer lies on the boundary. Suppose $D$ has an eigenvalue $d_i = 0$
with eigenvector $u_i$. Choose an eigenvalue $d_j > 0$ with eigenvector $u_j$;
one exists because $\operatorname{Tr} D = N_\sigma > 0$. The traceless
direction $X = u_i u_i^{\top} - u_j u_j^{\top}$ is feasible for
small $t > 0$. Its one-sided derivative
$\mathrm{d}/\mathrm{d}t\,\mathcal F_H(D + tX)|_{t = 0^{+}}$ equals
$\operatorname{Tr}(HX) + \tau \lim_{t \to 0^{+}} [h'(t) - h'(d_j - t)]$,
which is $-\infty$. Here $h'(t) = \ln(t/(1-t)) \to -\infty$, whereas
$-h'(d_j - t)$ is finite for $d_j < 1$ and also tends to $-\infty$ for
$d_j = 1$ because of the entropy-gradient singularity. The case $d_i = 1$
is symmetric: move weight to any
$d_j < 1$, which exists because $N_\sigma < n$. A feasible strict descent
direction therefore exists at every boundary point, even if other eigenvalues
remain on the boundary after this step. Every minimizer
is hence interior, $0 \prec D^\ast \prec I$, where the
strong-convexity estimate \eqref{eq:strong-convexity-general} applies;
and $4\tau$-strong convexity implies strict convexity, so the
minimizer is unique. The numerical evaluation therefore requires occupations
to lie in the open Fermi interval $(0,1)$.

\paragraph{Scope and extensions.}
The proof and the verified kernel are stated for real symmetric $D$
(matching the implementation, which diagonalizes a real symmetric
Hamiltonian through the dense self-adjoint path of
Section~\ref{sec:algorithms}) and for the unweighted Frobenius inner
product. Two extensions follow. (i) For complex Hermitian $D$, the same
divided-difference argument applies with $|\tilde X_{ij}|^{2}$ unchanged.
The implementation uses the faithful real-symmetric embedding
\begin{equation}
\mathcal{R}(A)=
\begin{pmatrix}
\operatorname{Re}A & -\operatorname{Im}A\\
\operatorname{Im}A & \operatorname{Re}A
\end{pmatrix},
\label{eq:real-embedding}
\end{equation}
which preserves the $4\tau$ constant and multiplies both the bound and
the directly evaluated distance by $\sqrt{2}$. (ii) Under the
matching-weight construction of
the global-$k$ screen, in which the functional
$\mathcal F = \sum_k w_k\, \mathcal F_k$ and the metric
$\langle X, Y \rangle_w = \sum_k w_k \operatorname{Tr}(X_k Y_k)$ carry
the same weights, the strong-convexity modulus remains $4\tau$. Applying the
divided-difference estimate block by block,
$\mathrm{D}^2\mathcal F[X,X] = \sum_k w_k\,
\mathrm{D}^2\mathcal F_k[X_k, X_k] \geq
4\tau \sum_k w_k \|X_k\|_F^2 = 4\tau \|X\|_w^2$,
the weights cancelling at every step, and the sharpness discussion above
is unaffected. More generally, if the functional carries positive weights
$a_k$ and the metric carries positive weights $b_k$, the modulus is
$4\tau\min_k(a_k/b_k)$. The numerical tests use $a_k=b_k=w_k$ and cover
both the unweighted
per-window case and the matching-weight global-k construction for graphene
($310$ trials) and fcc Al ($400$ trials).

\paragraph{Tangent residual and candidate bounds.}
Define the fixed-electron tangent residual
\begin{equation}
r(D) \;=\; \big\| \Pi_0 \big( H + \tau \, \ln\!\big( D (I-D)^{-1} \big) \big) \big\|_F ,
\qquad
\Pi_0 g = g - \frac{\operatorname{Tr} g}{n} \, I ,
\label{eq:tangent-residual}
\end{equation}
where the logarithm is spectral and $\Pi_0$ projects the Euclidean Mermin
gradient onto the traceless fixed-electron tangent space.
Strong convexity with modulus $\alpha = 4\tau$ yields the two candidate bounds
used by the screen: for any admissible $D$ and the canonical minimizer
$D^\ast$ of the same trace,
\begin{equation}
\| D - D^\ast \|_F \;\leq\; \frac{r(D)}{4\tau},
\qquad
\mathcal F_H(D) - \mathcal F_H(D^\ast) \;\leq\; \frac{r(D)^2}{8\tau}.
\label{eq:mermin-bounds}
\end{equation}
At the constrained minimizer, first-order optimality gives
\begin{equation*}
\nabla \mathcal F_H(D^\ast)=\mu I,
\qquad
\langle \nabla \mathcal F_H(D^\ast),D-D^\ast\rangle
=\mu\operatorname{Tr}(D-D^\ast)=0.
\end{equation*}
The standard strong-convexity inequalities
\cite[Sec.~9.1.2]{boyd2004convex} apply with the projected gradient
$\Pi_0 \nabla \mathcal F_H(D)$:
$\|D - D^\ast\|_F \leq
\|\Pi_0 \nabla \mathcal F_H(D)\|_F / \alpha$ (gradient
monotonicity against the minimizer) and
$\mathcal F_H(D) - \mathcal F_H(D^\ast) \leq
\|\Pi_0 \nabla \mathcal F_H(D)\|_F^2 / (2\alpha)$.
The first bound limits how far a candidate density can be from the
canonical one; the second bounds the free-energy suboptimality.

For the global-$k$ construction, these inequalities hold on the exact
weighted-trace affine space
$\sum_k w_k\operatorname{Tr}D_{\sigma,k}=N_\sigma$.
The mathematical embedding satisfies
$\operatorname{Tr}\mathcal R(D_{\sigma,k})=2\operatorname{Tr}D_{\sigma,k}$.
Separately, physical spin degeneracy gives $N_e=2N_\sigma$.
Thus the embedded per-spin trace equals $N_e$ in this representation;
embedding and spin degeneracy are distinct operations, not two successive
factors multiplying the electron number.
The historical controller first applied a separate trace gate;
comparisons with computed reference distances used an empirical
floating-point allowance.
A nonzero trace tolerance is a numerical acceptance condition,
not an extension of the exact-trace theorem. The empirical allowance and its
limitations are stated in Sec.~\ref{sec:controller-tests}.

\paragraph{Particle-number defect correction.}
The distance bound admits a separate correction when the candidate trace
differs from the target. For one block, let $M=\operatorname{Tr}D$ and
let $D_M^\star$ and $D_N^\star$ be the canonical states of the same
fixed Hamiltonian at traces $M$ and $N$, respectively. Assume
$0\prec D\prec I$, $\tau>0$ and $0<N<n$.
Applying Eq.~\eqref{eq:mermin-bounds} at trace $M$ gives
$\|D-D_M^\star\|_F\le r(D)/(4\tau)$.
Every Fermi occupation increases with the common chemical potential.
Consequently $D_M^\star-D_N^\star$ has a common spectral sign, and
\begin{equation}
 \|D_M^\star-D_N^\star\|_F
 \le\operatorname{Tr}|D_M^\star-D_N^\star|=|M-N|.
\end{equation}
The triangle inequality therefore yields
\begin{equation}
 \boxed{\|D-D_N^\star\|_F\le\frac{r(D)}{4\tau}
                  +|\operatorname{Tr}D-N|.}
 \label{eq:trace-defect-distance}
\end{equation}

For positive weights, write
$\delta_\sigma=\sum_k w_k\operatorname{Tr}D_k-N_\sigma$,
$w_{\min}=\min_k w_k>0$, and let $r_w$ be the residual in the
matching weighted metric. The target must lie strictly between zero
and the weighted state capacity $\sum_k w_k n_k$.
The same argument gives
\begin{equation}
 \|D-D^\star\|_w\le\frac{r_w(D)}{4\tau}
              +\frac{|\delta_\sigma|}{\sqrt{w_{\min}}}.
 \label{eq:weighted-trace-defect}
\end{equation}
Indeed, for the difference $A_k$ between canonical states at the two
weighted traces, set $t_k=\operatorname{Tr}|A_k|$. Their common sign implies
$\sum_k w_k t_k=|\delta_\sigma|$, while
\begin{equation}
 \sum_k w_k\|A_k\|_F^2\le\sum_k w_k t_k^2
 \le\frac{(\sum_k w_k t_k)^2}{w_{\min}}.
\end{equation}
In the spin-degenerate real embedding, the norm is multiplied by $\sqrt2$
and $\delta N_e=2\delta_\sigma$. The additive distance correction is then
$|\delta N_e|/\sqrt{2w_{\min}}$.
For a retained window, the target is its prescribed electron number after
subtracting the frozen complement. The subspaces, complement, Hamiltonian,
weights and temperature remain fixed during this comparison.
These coefficients are sharp over Hamiltonians: an isolated thermally
active level can concentrate the trace change in the least-weighted block.
The other occupation changes approach zero as their energies move away
from the chemical potential.

Independent finite-matrix checks used 400 seeded trace-shift and unitary-rotation
constructions, with one to four weighted blocks and temperatures
$0.03\le\tau\le0.5$~Ha. No corrected distance bound was exceeded by more
than $10^{-10}$; the largest distance-to-bound ratio was $0.879$.
The uncorrected expression was exceeded in 189 trace-defective inputs,
which lie outside its exact-trace hypothesis.
Two canonical sequences evaluated at 80 decimal digits approach the
single-block and weighted correction constants within $10^{-9}$.
This corollary controls particle-number mismatch in exact arithmetic.
It does not bound errors in evaluating the residual, extend a window
estimate to the full density, or correct the free-energy inequality for
unequal traces. The historical controller calculations predate this
correction and are not reclassified by it. The separate historical-subspace
tests in Sec.~\ref{sec:historical-materials} use the corrected distance bound.

\paragraph{Numerical corroboration.}
An independent arbitrary-precision implementation, using its own eigensolver and bisection
chemical-potential solve at 50 and 100 decimal digits, also recomputed the
complete chain. The Loewner-Hessian identity of
Eq.~\eqref{eq:loewner-hessian} agreed with Richardson-extrapolated finite
differences to a largest relative difference of $1.93\times10^{-64}$.
The $10{,}000$-sample set gave no violation above
$1+10^{-30}$, including the complex-Hermitian and weighted-$k$ extensions,
and the analytic chemical potential was recovered within
$1.91\times10^{-90}$. The largest relative difference between the
double-precision implementation and an independent double-precision calculation
was $6.9\times10^{-7}$. These calculations test the implementation
independently of the analytic proof; they are not interval enclosures.

\subsection{Why a zero window residual does not bound the full-space error}
\label{sec:window-counterexamples}
Let $P$ have orthonormal columns and let the screened reference be the
canonical minimizer for $P^\dagger HP$ at the prescribed window trace.
It need not equal $P^\dagger D^\ast P$, where $D^\ast$ is the canonical
state of the full Hamiltonian. Two finite-dimensional examples separate
the missing complement and coupling information. All densities below are
per-spin matrices in the ordinary, unembedded Frobenius metric.

First, take $\tau=0.05$~Ha,
\begin{equation}
H=\operatorname{diag}(0,0,-1,1)\ \mathrm{Ha},\qquad
P=(e_1,e_2),\qquad
D=\operatorname{diag}(1/2,1/2,0.05,0.95).
\end{equation}
Both $D$ and the full canonical state have trace two. With
$p=(1+e^{-20})^{-1}$, the latter is
$D^\ast=\operatorname{diag}(1/2,1/2,p,1-p)$.
The window density is exactly canonical, so its residual and error vanish;
the Hamiltonian coupling to the complement also vanishes. Nevertheless,
\begin{equation}
\|D-D^\ast\|_F=\sqrt{2}(p-0.05)\simeq1.344.
\end{equation}

Second, take $\tau=0.05$~Ha, $t=0.5$~Ha and
\begin{equation}
H=\begin{pmatrix}0&t\\t&0\end{pmatrix},\qquad D=I/2,
\qquad \operatorname{Tr}D=1.
\end{equation}
Each one-dimensional compression has zero Hamiltonian and the exact
occupation $1/2$. The full canonical density instead has off-diagonal
entries $-\tfrac12\tanh[t/(2\tau)]$, giving
\begin{equation}
\|D-D^\ast\|_F=\frac{\tanh[t/(2\tau)]}{\sqrt{2}}
\simeq0.707.
\end{equation}
Thus no universal finite constant can bound the full-space error by the
window residual alone without additional complement and coupling
conditions. These examples are not material calculations or demonstrations
of false acceptance by the complete controller: the frozen-edge, charge,
and minimum-window-dimension checks impose additional restrictions.

\paragraph{Quantifying the omitted coupling.}
The same strong-convexity estimate gives a constructive separation of these
two error sources. Put $R=PP^\dagger$, $Q=I-R$, and
$H_0=RHR+QHQ$. Let $D_H^\ast$ and $D_0^\ast$ be the full canonical
states of $H$ and $H_0$ at the same $N_\sigma$ and $\tau$.
In particular, the blocks of $D_0^\ast$ share one chemical potential;
their particle numbers are not fixed independently.
With $\Delta D=D_0^\ast-D_H^\ast$, strong monotonicity and stationarity give
\begin{equation}
\begin{aligned}
4\tau\|\Delta D\|_F^2
&\le \langle \nabla\mathcal F_H(D_0^\ast)
                 -\nabla\mathcal F_H(D_H^\ast),\Delta D\rangle_F\\
&=\langle H-H_0,\Delta D\rangle_F\\
&\le \|\Pi_0(H-H_0)\|_F\|\Delta D\|_F .
\end{aligned}
\end{equation}
The chemical-potential terms vanish because $\operatorname{Tr}\Delta D=0$.
The off-diagonal perturbation has zero trace and squared norm
$2\|QHP\|_F^2$. Thus
\begin{equation}
\|D_H^\ast-D_0^\ast\|_F
\le \frac{\sqrt{2}\|QHP\|_F}{4\tau}.
\label{eq:block-coupling}
\end{equation}
For any block-diagonal candidate $D=RDR+QDQ$, define
$e_P=\|P^\dagger(D-D_0^\ast)P\|_F$ and
$e_Q=\|Q(D-D_0^\ast)Q\|_F$. Orthogonality and the triangle inequality yield
\begin{equation}
\|D-D_H^\ast\|_F
\le \sqrt{e_P^2+e_Q^2}
       +\frac{\sqrt{2}\|QHP\|_F}{4\tau}.
\label{eq:full-block-decomposition}
\end{equation}
The first term includes complement occupations and particle allocation
between blocks. It cannot be replaced by the residual at a prescribed
window trace. Evaluating $D_0^\ast$ generally requires complement information,
so this decomposition does not itself supply an inexpensive full-state test.
It remains valid for endpoint occupations in $D$, since no logarithm of
the candidate is used.

For weighted k points, the same proof uses the weighted inner product
and one total trace constraint; $e_P^2$ and $e_Q^2$ become the corresponding
weighted sums of squared block errors. The coupling term becomes
$[2\sum_k w_k\|Q_kH_kP_k\|_F^2]^{1/2}/(4\tau)$.
More generally, a Hamiltonian perturbation is bounded by its norm after
removing one common scalar, not separate shifts at each k point.
The spin-degenerate real embedding multiplies both distances and the
coupling term by $\sqrt{2}$.

An independent test generated 400 finite-matrix cases, divided equally
among real and complex single-block and weighted-k constructions.
Dimensions were 3--9, with one to four k blocks and
$0.02\le\tau\le0.5$~Ha. The coupling and full-distance estimates had no
violations above $10^{-10}$; their maximum ratios were $0.956$ and $0.994$.
Reference and candidate particle-number errors remained below
$1.750\times10^{-13}$. In the two-level example, the analytic ratio
$2\tanh[t/(2\tau)]/(t/\tau)$ tends to unity as $t/\tau\to0$,
so the coupling coefficient is sharp. Four 80-digit evaluations
corroborate this limit without attaining equality at nonzero coupling.
These tests concern fixed finite matrices, not material SCF trajectories.

\section{Numerical verification of the finite-temperature bounds}
\label{sec:numerical-verification}

\paragraph{Randomized-matrix test.}
The finite-temperature reference kernel is verified on $10{,}000$ seeded
random real symmetric matrices: dimensions $2$--$8$, seeded SplitMix64
generation, and three ensembles chosen to stress the bound regimes --
near-degenerate spectra, near-saturated occupations and half occupation.
Each sample checks the exact first Fr\'echet derivative against central
finite differences (observed order), the trace constraint, and both
Mermin residual bounds of Eq.~\eqref{eq:mermin-bounds} against a
same-trace trial density; any violation fails the test.

\paragraph{Results.}
No bound violation occurs in $10{,}000$ samples. The median
distance-to-bound ratio is $0.632$, with a maximum printed as $1.000$ at
half occupation. The median free-energy ratio is $0.614$. The median
central-difference order is $2.000$ (minimum $1.92$), and the largest trace
error is $1.65\times10^{-12}$.

\paragraph{Eigensolver sensitivity.}
A near-degenerate test exposed inaccurate eigenvectors from an earlier
self-adjoint solver: reconstruction residuals were $8.9\times10^{-4}$ at a
cluster gap of $10^{-4}$ and $5.9\times10^{-2}$ at a gap of $10^{-5}$,
although the eigenvalues agreed to $10^{-14}$. The corrected reference path
uses a safeguarded self-adjoint solver with orthogonality and per-eigenpair
relative-residual criteria of $10^{-11}$.

\paragraph{Common-energy offsets and finite precision.}
The exact tangent residual is unchanged by $H\mapsto H+cI$.
We tested this property directly in the real, complex, weighted-real and
weighted-complex numerical kernels with
\begin{equation}
H=cI_4,\qquad \tau=1~\mathrm{Ha},\qquad
D=\operatorname{diag}(0.7,0.5,0.7,0.5).
\label{eq:offset-test}
\end{equation}
The target trace is $2.4$, the canonical reference is $D^\ast=0.6I_4$,
and the unembedded Frobenius distance is $0.200$.
For a single k block of unit weight, all four kernels give
$r=0.847$~Ha and $r/(4\tau)=0.212$ at $c=0$.
At $c=10^6$~Ha, the residual changes by $3.869\times10^{-11}$~Ha.
In the earlier implementation, all four uncentered evaluations returned zero
at $c=10^{16}$~Ha,
which cannot bound the nonzero distance in Eq.~\eqref{eq:offset-test}.

The cancellation occurs when the entropy gradient is added to the large
constant Hamiltonian before its scalar part is removed.
Evaluating the same kernels with $H-cI$ restores $r=0.847$~Ha.
The algebraically equivalent form
\begin{equation}
G_T=\Pi_0H+\tau\Pi_0\ln[D(I-D)^{-1}]
\end{equation}
avoids this particular loss of the entropy contribution.
For weighted k points, the removed energy offset must be common to all
blocks; independent block shifts alter their relative occupations.
These observations identify a floating-point implementation failure,
not a counterexample to the exact-trace inequality.

The reference solve also requires an explicit trace check.
At $c=10^{16}$~Ha, the earlier scalar canonical routine returned a nominally
successful state with trace $2.000$ instead of $2.400$.
The shared-chemical-potential routine reports failure for the same input.
At $c=10^6$~Ha, the scalar trace error is $2.249\times10^{-10}$,
and the shared routine again reports failure at its stricter tolerance.
The analytic state, rather than either failed numerical reference,
defines the distance in this test.

Direct evaluation of the screening decision with the corresponding
spin-degenerate proposal gives a rejection at $c=0$ and $10^6$~Ha.
At $c=10^{16}$~Ha the earlier implementation accepted a zero bound, although the analytic embedded
distance is $\sqrt{2}(0.200)=0.283$, above the $0.05$ threshold.
The numerical reference distance is unavailable at this offset.
This was a failure of the supplied-matrix screening decision; no material
SCF trajectory was propagated in this test.

The corrected implementation removes a common energy origin before adding
the entropy gradient when the uncentered sum cannot resolve that contribution.
The scalar canonical solve evaluates occupations in relative-energy coordinates
at extreme offsets and checks the resulting trace before returning a state.
The four residual variants recover $r=0.847$~Ha at $c=10^{16}$~Ha;
the scalar reference has trace $2.400$, and the embedded screening example
is rejected with a distance bound of $0.300$.
Unresolvable scales return an explicit error.
These corrections were checked in the release build used for the
historical-subspace material comparisons described below.

\paragraph{Reference-failure states.}
Synthetic endpoint pairs were passed directly to the terminal comparison
and decision routines, without running an SCF trajectory.
The earlier routine detected nonconvergence of the controlled endpoint,
but did not report a failure when only the reference was unconverged.
Likewise, NaN free-energy or density differences, and the undefined
difference of two infinite free energies, yielded no reported failure.
These outcomes require explicit reference-convergence and finiteness checks;
threshold comparisons alone do not implement those conditions.
The corrected routine requires both endpoints to be converged and every
compared quantity to be finite. All four defective endpoint pairs are
rejected in direct regression tests; a finite, converged matching pair passes.
The separate density-payload guard rejects nonfinite values, negative
density and excessive charge error in its direct tests.

\paragraph{Discretization checks.}
Separate finite-difference response checks were applied to a cutoff-by-grid
factorial (nine cells, all response criteria satisfied, finest-edge
Jacobian changes $2.55\times10^{-13}$ and $6.81\times10^{-8}$) and the
k-point nested-mesh calculation (no response-criterion violations).
These test SCF response derivatives, not sharpness of the fixed-$H$ bound.
The detailed per-level tables of these two checks are reported in
Tables~\ref{tab:mg-factorial-rungs} and~\ref{tab:k2-nested-rungs}.

\begin{table}[tbp]
\centering
\caption{Cutoff-by-grid factorial: nine
(cutoff, grid) cells at fixed coupling ladder $\lambda \in \{0, 0.2,
0.4\}\,\lambda_c$ with $\lambda_c = -47.918$~Ha\,bohr$^3$, the projected critical
coupling of the three-mode $\Gamma$-point plane-wave model (not of the
real-space H$_2$/Grid4 model, whose $\lambda_c = -59.076$~Ha\,bohr$^3$ belongs to
a different reduced SCF map). Every cell satisfies the numerical tests.
The finest-cutoff edge ($0.70 \to 1.05$~Ha at $12^3$) changes
the projected Jacobian by $2.55\times 10^{-13}$ (relative) and
the finest-grid edge ($10^3 \to 12^3$ at $1.05$~Ha) by
$6.81\times 10^{-8}$; subspace overlap is unity to the displayed precision on both
edges.}
\label{tab:mg-factorial-rungs}
\begin{tabular}{cccccc}
\toprule
cutoff (Ha) & grid & basis & $m_{\mathrm{fp}}(0)$ &
$m_{\mathrm{fp}}(0.4\,\lambda_c)$ & energy-curvature error \\
\midrule
$0.35$ & $8^3$  & $7$  & $1.000$ & $0.702$ & $1.24 \times 10^{-8}$ \\
$0.35$ & $10^3$ & $7$  & $1.000$ & $0.702$ & $1.41 \times 10^{-8}$ \\
$0.35$ & $12^3$ & $7$  & $1.000$ & $0.702$ & $1.47 \times 10^{-8}$ \\
$0.70$ & $8^3$  & $19$ & $1.000$ & $0.702$ & $1.25 \times 10^{-8}$ \\
$0.70$ & $10^3$ & $19$ & $1.000$ & $0.702$ & $1.45 \times 10^{-8}$ \\
$0.70$ & $12^3$ & $19$ & $1.000$ & $0.702$ & $1.57 \times 10^{-8}$ \\
$1.05$ & $8^3$  & $27$ & $1.000$ & $0.702$ & $1.32 \times 10^{-8}$ \\
$1.05$ & $10^3$ & $27$ & $1.000$ & $0.702$ & $1.41 \times 10^{-8}$ \\
$1.05$ & $12^3$ & $27$ & $1.000$ & $0.702$ & $1.40 \times 10^{-8}$ \\
\bottomrule
\end{tabular}
\end{table}

\begin{table}[tbp]
\centering
\caption{Nested-mesh check: per-mesh fixed-coupling levels at
$\lambda \in \{0, 0.2, 0.4\}\,\lambda_c(\mathrm{mesh})$ in the weighted-k
reduced SCF model, whose $\lambda_c$ is defined per
mesh (the two finest meshes already agree to $1.3\%$; neither value is
comparable to the $\Gamma$-point plane-wave or real-space couplings above). All three
meshes satisfy the numerical tests. Mesh edges: $1^3 \to 2^3$
(relative Jacobian
change $1.859 \times 10^{-1}$) and $2^3 \to 4^3$ (the checked refinement
edge: relative Jacobian change $1.832 \times 10^{-2}$, susceptibility
change $1.464 \times 10^{-2}$, margin changes $2.1 \times 10^{-3}$ and
$5.2 \times 10^{-3}$, soft-residual-subspace overlap
$1.000$ to the displayed precision).}
\label{tab:k2-nested-rungs}
\begin{tabular}{ccccccc}
\toprule
mesh & $k$-points & basis & $\lambda_c$ (Ha\,bohr$^3$) & $m_{\mathrm{fp}}(0)$ &
$m_{\mathrm{fp}}(0.4\,\lambda_c)$ & $\chi(0.4\,\lambda_c)/\chi(0)$ \\
\midrule
$1^3$ & $1$  & $19$      & $-42.013$ & $0.963$ & $0.792$ & $1.667$ \\
$2^3$ & $8$  & $17$      & $-36.311$ & $0.987$ & $0.833$ & $1.667$ \\
$4^3$ & $64$ & $11$--$16$ & $-36.786$ & $0.990$ & $0.828$ & $1.667$ \\
\bottomrule
\end{tabular}
\end{table}

\section{H\(_2\) toy model derivations}
\label{sec:h2toy}

The toy substrate is a closed-shell H$_2$/Grid4 real-space model with a
softened local Coulomb potential, used to test the distinction between
intrinsic response and mixer amplification at a scale where every quantity can be recomputed
exactly. The calculation includes the exact charge-neutral SCF
Jacobian at fixed electron number; a fixed-point-preserving rank-one
response intervention evaluated at three intervention strengths against
four fixed mixers; an independently reconverged external-field
susceptibility as the intervention-free reference; and a directional
Grid4-to-Grid6 refinement check.

\paragraph{Rank-one, fixed-point-preserving response intervention.}
All objects live on the discrete real-space grid with the weighted
inner product
$\langle f, g \rangle_W = \Delta\mathcal{V}
\sum_{\mathbf{r}} f(\mathbf{r})\, g(\mathbf{r})$
($\Delta\mathcal{V}$ is the grid volume element) and its induced norm.
The self-consistent map is
$M(\rho) = \Phi\!\big(V_{\mathrm{eff}}[\rho]\big)$, where
$V_{\mathrm{eff}}[\rho] = V_{\mathrm{ext}} + V_{\mathrm{H}}[\rho] + V_{\mathrm{xc}}[\rho]$
and $\Phi$ solves the band problem at fixed electron number and
returns the density (a constant potential shift is removed as a
gauge before diagonalization).
Fix a converged anchor $\rho^\ast = M(\rho^\ast)$ and a charge-neutral
mode $v$ with $\langle v, v \rangle_W = 1$; the check uses the lowest
cosine mode of the cell, mean-subtracted and normalized.
With the physical density measured in bohr$^{-3}$, this normalization
gives $\lambda$ units of Ha\,bohr$^{3}$, so that
Eq.~\eqref{eq:intervention-energy} has energy units and
Eq.~\eqref{eq:intervention-potential} has potential units.
The intervention is generated at the energy level by adding the
anchor-centered quadratic term
\begin{equation}
\Delta E_\lambda[\rho] \;=\; \frac{\lambda}{2}\,
\langle v, \rho - \rho^\ast \rangle_W^{2},
\label{eq:intervention-energy}
\end{equation}
whose first variation shifts the effective potential by
\begin{equation}
\Delta V_\lambda[\rho] \;=\; \frac{\delta \Delta E_\lambda}{\delta \rho}
\;=\; \lambda\, v\, \langle v, \rho - \rho^\ast \rangle_W ,
\label{eq:intervention-potential}
\end{equation}
so the intervened fixed-point map is
\begin{equation}
M_\lambda(\rho) \;=\; \Phi\!\big(V_{\mathrm{eff}}[\rho] + \Delta V_\lambda[\rho]\big).
\label{eq:intervened-map}
\end{equation}

\paragraph{Fixed-point preservation (variational argument).}
The amplitude $\langle v, \rho - \rho^\ast \rangle_W$ vanishes at
$\rho = \rho^\ast$, so both the added energy \eqref{eq:intervention-energy}
and its first variation \eqref{eq:intervention-potential} vanish at the
anchor: $\Delta V_\lambda[\rho^\ast] = 0$ and therefore
$M_\lambda(\rho^\ast) = M(\rho^\ast) = \rho^\ast$ for every $\lambda$.
The intervention is algebraically zero at the anchor, and the anchor
remains a fixed point of the intervened map with the same output
density. The numerical anchor is converged to a finite tolerance. An
independent zero-field reconvergence therefore verifies recovery of the same
density, potential, and free energy and excludes a fixed-point shift from the
residual anchor error.

\paragraph{Linearization: exact rank-one Jacobian update.}
Let $\chi = D\Phi|_{V_{\mathrm{eff}}[\rho^\ast]}$ be the
density-potential response (independent susceptibility) at the anchor,
measured by central differences of $\Phi$ without any reconvergence,
and let $K = DV_{\mathrm{eff}}|_{\rho^\ast}$ be the Hartree--xc kernel,
so the neutral SCF Jacobian is $J = \chi K$. The chain from the energy
term \eqref{eq:intervention-energy} through the potential
\eqref{eq:intervention-potential} to the map \eqref{eq:intervened-map}
gives
$D\Delta V_\lambda|_{\rho^\ast}[\delta\rho] = \lambda\, v\, \langle v, \delta\rho \rangle_W$
and hence
\begin{equation}
J_\lambda \;=\; J \;+\; \lambda\, u\, \langle v, \cdot \rangle_W,
\qquad u \;=\; \chi v ,
\label{eq:rank-one-jacobian}
\end{equation}
an exact rank-one update: $\chi$ and $v$ are held fixed at the anchor, so
there is no higher-order remainder in $\lambda$. The same linearization
predicts the intervened directional response of the map along the mode
itself as $Jv + \lambda\, \chi v$ (using $\langle v,v\rangle_W = 1$),
which is checked against central differences of the intervened
map.

\paragraph{Critical strength from the determinant lemma.}
Write $A = I - J$ for the fixed-point operator and
$A_\lambda = I - J_\lambda = A - \lambda\, u \langle v, \cdot \rangle_W$.
Assuming $A$ is invertible, the matrix determinant lemma reduces the
singularity condition of a rank-one perturbation to one scalar projection:
$A_\lambda$ is singular exactly when
$1 - \lambda\, \langle v, A^{-1} u \rangle_W = 0$, i.e.
\begin{equation}
\lambda_c \;=\; \frac{1}{\langle v, A^{-1} u \rangle_W}
\;=\; \frac{1}{\langle v, \chi_{\mathrm{sc}}\, v \rangle_W},
\qquad
\chi_{\mathrm{sc}} \;=\; (I - J)^{-1} \chi ,
\label{eq:critical-lambda}
\end{equation}
where $\chi_{\mathrm{sc}}\, v = A^{-1} u$ is the linearized
self-consistent susceptibility along $v$. An independent estimate is
obtained by reconverging the fixed point under an external field
$\pm h v$ and taking a central difference. The check evaluates the same scalar in
the reduced charge-neutral basis by solving $(I - J)\, x = \chi v$
directly and confirms that the predicted critical residual operator
$I-J-\lambda_c\,u\langle v,\cdot\rangle_W$ is singular to working
precision.

\paragraph{Intervened susceptibility: Sherman--Morrison scaling.}
When the probe field is aligned with the intervention mode, $A$ is
invertible, and
$1-\lambda\langle v,A^{-1}u\rangle_W\neq0$, the response source is
$u = \chi v$ itself, and the Sherman--Morrison
formula applied to $A_\lambda = A - \lambda\, u \langle v, \cdot \rangle_W$
gives the intervened self-consistent susceptibility in closed form,
\begin{equation}
\chi_{\mathrm{sc}}^{(\lambda)}\, v
\;=\; A_\lambda^{-1} u
\;=\; \frac{A^{-1} u}{1 - \lambda\, \langle v, A^{-1} u \rangle_W}
\;=\; \frac{\chi_{\mathrm{sc}}\, v}{1 - \lambda / \lambda_c} ,
\label{eq:sherman-morrison}
\end{equation}
so the entire response vector rescales by the single scalar
$1 / (1 - \lambda\, \langle v, \chi_{\mathrm{sc}} v \rangle_W)$,
diverging as $\lambda \to \lambda_c$. The check compares this
prediction against independently reconverged intervened field
responses, and the structural rank-one identity
\eqref{eq:rank-one-jacobian} is checked at three intervention
strengths $\lambda \in \{0.10, 0.20, 0.40\}\,\lambda_c$ against four
fixed preconditioned mixers.

\paragraph{Scope of the intervention.}
The construction controls the specified anchor
fixed point and its linearized response. Equations
\eqref{eq:intervention-energy}--\eqref{eq:sherman-morrison} guarantee
that $\rho^\ast$ is preserved and that its Jacobian and mode-aligned
susceptibility move in the predicted rank-one way. The added term
\eqref{eq:intervention-energy} is an anchor-centered quadratic with
zero value and zero first variation at the anchor. Global energy surfaces,
additional fixed points, and convergence basins are distinct properties and
are not used in the causal identification.

\begin{figure}[tbp]
\centering
\includegraphics[width=\textwidth]{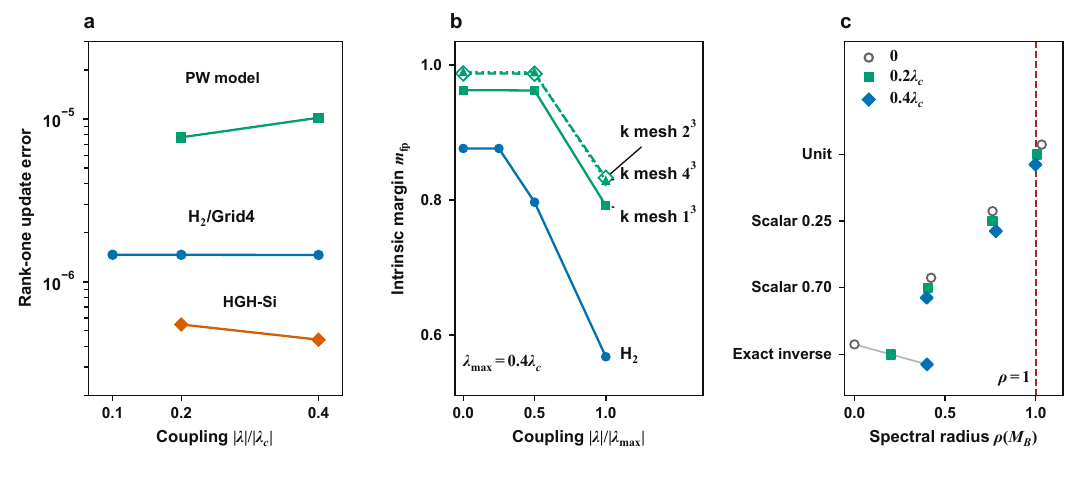}
\caption{Fixed-point-preserving interventions and mixer response.
\textbf{(a)} Rank-one Jacobian-update errors versus
$|\lambda|/|\lambda_c|$ for the real-space H$_2$/Grid4 model,
the three-mode $\Gamma$-point plane-wave model, and HGH silicon.
Markers show numerical diagnostics; lines connect the tested points.
\textbf{(b)} Intrinsic margin $m_{\mathrm{fp}}$ for H$_2$/Grid4 and the
weighted-k reduced model on $1^3$, $2^3$ and $4^3$ meshes.
The abscissa is $|\lambda|/|\lambda_{\max}|$, with
$\lambda_{\max}=0.4\lambda_c$ separately for each model and mesh;
unity is not the critical coupling or a common absolute coupling.
\textbf{(c)} Spectral radius $\rho(M_B)$ for four mixers in HGH silicon.
Open circles, squares and diamonds denote $0$, $0.2\lambda_c$ and
$0.4\lambda_c$, respectively; vertical offsets separate strengths within
each category. The dashed line marks $\rho(M_B)=1$.
The exact inverse of the baseline residual operator is an algebraic control.
The Si intrinsic margin changes from $0.944$ to $0.854$ between baseline
and $0.4\lambda_c$. These local response checks are independent of the
strong-convexity proof and do not establish global convergence.
AI-assisted figure preparation was checked against the numerical records
and mathematical definitions.}
\label{fig:si-intervention}
\end{figure}

For a fixed preconditioner $B$, the linearized mixed iteration is
$M_B=I-B(I-J)$. Changing $B$ at fixed $J$ changes its spectral radius
and transient amplification without changing the intrinsic margin
$m_{\mathrm{fp}}=\sigma_{\min}(I-J)$. Figure~\ref{fig:si-intervention}
summarizes the intervention and mixer tests; numerical details are in
Sec.~\ref{sec:intervention-results}. These response tests are independent
of the strong-convexity proof in Sec.~\ref{sec:mermin}.

\section{Numerical algorithms}
\label{sec:algorithms}

OpenAI Codex provided substantive AI assistance with coding, analysis, and
manuscript preparation; analytical checks, stored numerical outputs, and
source checks were used to assess the resulting code and scientific claims.

\subsection{Kleinman--Bylander nonlocal operator assembly}
The implementation uses the separable Kleinman--Bylander form
\cite{kleinman1982}.
Before channel diagonalization, the separable nonlocal operator is
\begin{equation}
V_{\mathrm{NL}}=
\sum_{a,l,m}\sum_{i,j}h_{ij}^{l}
|\beta^{i}_{alm}\rangle\langle\beta^{j}_{alm}|.
\label{eq:nonlocal-general}
\end{equation}
For each $l$, the implementation diagonalizes
$h^{l}=U^{l}\operatorname{diag}(d_{l\alpha})(U^{l})^{\top}$ and rotates the
radial projectors,
$q_{l\alpha}=\sum_i U^{l}_{i\alpha}\beta^i_l$. Thus
\begin{equation}
V_{\mathrm{NL}}=
\sum_{a,l,m,\alpha}d_{l\alpha}
|\tilde q_{a l m\alpha}\rangle
\langle\tilde q_{a l m\alpha}|,
\label{eq:nonlocal-diagonal}
\end{equation}
where
$\tilde q_{a l m\alpha}(\mathbf r)=
q_{l\alpha}(|\mathbf r-\boldsymbol{\tau}_a|)
Y_{lm}(\widehat{\mathbf r-\boldsymbol{\tau}_a})$.
Plane waves are normalized as $e^{i(\mathbf{k}+\mathbf{G})\cdot\mathbf{r}}/\sqrt{\Omega}$
(Bohr/Hartree atomic units throughout). With
$\widehat q_{l\alpha}(q) = \int r^2 j_l(qr)\,
q_{l\alpha}(r)\,\mathrm{d}r$ the
matrix element is
$\langle \mathbf{k}+\mathbf{G} | V_{\mathrm{NL}} | \mathbf{k}+\mathbf{G}' \rangle
= \sum_{a,l,m,\alpha}d_{l\alpha}\Omega\,
\overline{b_{a l m\alpha}(\mathbf{k}+\mathbf{G})}\,
b_{a l m\alpha}(\mathbf{k}+\mathbf{G}')$ with
$b_{a l m\alpha}(\mathbf{q}) = (4\pi/\Omega)\, i^l\,
Y_{lm}(\hat{\mathbf{q}})\,
\widehat q_{l\alpha}(q)\,
e^{i\mathbf{q}\cdot\boldsymbol{\tau}_a}$;
the prefactor, the $i^l$ phase and the structure-factor sign were checked
against direct real-space integration. The $Y_{lm}$ are real orthonormal spherical
harmonics ordered by $m=-l,\ldots,l$, implemented for $l \leq 2$; $l = 3$
channels are parsed but are not evaluated in the present calculations.

\subsection{Block Davidson eigensolver with a safeguarded residual criterion}
All requested bands are corrected every iteration (block variant), which
treats exactly degenerate plane-wave kinetic shells without per-vector
locking. The Jacobi preconditioner $(\theta - \operatorname{diag} H)^{-1}$
uses a sign-preserving floor of $10^{-3}$ in matrix units; the trial
subspace is Rayleigh--Ritz solved each iteration with the same dense
\textsc{faer} kernel as the reference path; thick restart keeps the lowest Ritz
vectors whenever the subspace cap is reached; orthogonalization is
two-pass Gram--Schmidt with a $10^{-10}$ linear-dependence drop test, and
a round with no surviving correction direction is a stagnation error,
never a silent exit. Before returning, every requested eigenpair is
re-screened directly against the input operator:
$\|Hv - \varepsilon v\|_2/s_H\leq\varepsilon_{\mathrm{eig}}$.
Here $s_H=\max(\|H\|_F,1)$ for dense matrices and is a certified upper
bound on $\|H\|_2$ for matrix-free operators. The default
$\varepsilon_{\mathrm{eig}}$ is $10^{-11}$, and the returned set must be
orthonormal to $10^{-11}$. Any criterion
violation, invalid configuration, non-Hermitian input, iteration
exhaustion or stagnation terminates the eigensolve.

\subsection{Dense reference eigensolver path}
A dense self-adjoint \textsc{faer} path serves as the reference for
small systems. It aborts the solve on non-finite input, on orthonormality
deviation above $10^{-11}$, or on any per-column relative residual above
threshold. This path replaced the nalgebra symmetric eigensolver after the
defect documented in Section~\ref{sec:numerical-verification}.

\subsection{Canonical occupations and occupation clamping}
Occupations follow one shared canonical Fermi chemical potential solving
\begin{equation*}
N=\sum_i 2w_i f\!\left(\frac{\varepsilon_i-\mu}{\tau}\right)
\end{equation*}
over weighted states,
with occupations in $[0,2]$ per spatial state; state weights are
integration weights, not degeneracies. Convergence additionally requires
the highest retained occupation and the weighted edge-electron count to
satisfy a declared tolerance, so a converged density cannot hide an
under-resolved finite-temperature band tail. A $10^{-14}$ occupation
clamp is used for the density reconstruction of deeply occupied states.

\section{Numerical parameters}
\label{sec:numerical-parameters}

Table~\ref{tab:numerical-parameters} collects the numerical choices used in the
screening, response and grouped work-prediction calculations.

\newpage
\begin{table}[t]
\centering
\caption{Numerical parameters used in the reported calculations.}
\label{tab:numerical-parameters}
\begin{tabular}{@{}p{0.43\textwidth}p{0.49\textwidth}@{}}
\toprule
Item & Value \\
\midrule
Mixer settings, reduced models & linear $0.25$; secant--Kerker $0.70$, $q_0=0.30$, history 4 \\
Mixer arms, pseudopotential models & linear $0.25$ and $0.70$ \\
Default SCF iteration limit & 150 \\
Default Davidson relative-residual criterion & $10^{-11}$ \\
Dense-solver orthonormality criterion & $10^{-11}$ \\
Jacobi preconditioner floor & $10^{-3}$ in matrix units \\
Gram--Schmidt dependence threshold & $10^{-10}$ \\
Screening distance tolerance & $0.05$ (absolute Frobenius distance) \\
Initial window-count half-width & $18\tau$; common-$m$ selection may extend beyond it \\
Trace tolerance & $10^{-8}$ (reported kernel tests, caller supplied); $10^{-6}$ (controller) \\
Frozen outside-edge occupation tolerance & $10^{-6}$ (per spin) \\
Empirical numerical allowance & Eq.~\eqref{eq:screen-slack} \\
Terminal comparison tolerances & $10^{-7}$~Ha (free energy); $10^{-5}$ (density); $10^{-5}$~Ha (HOMO/LUMO) \\
Bootstrap RNG seeds & $\{13,29,47,73,101\}$; deterministic OLS fits \\
Inclusion rule & $A_k\leq0.85A^\ast$ \\
$A^\ast$ values & $0.30$ (reduced model); $0.023$ and $0.165$ (representative pseudopotential categories) \\
Uncertainty calculation & hierarchical bootstrap, $10^4$ replicates, 95\% interval \\
Cross-category reversal threshold & $-5\%$ relative error reduction \\
Minimum included-step coverage & $20\%$ \\
\bottomrule
\end{tabular}
\end{table}

AB-stacked graphene and the bilayer, bulk 2H and 1T$'$ MoS$_2$ families
use mixing $0.1$ and an iteration limit of $300$. AB-stacked graphene uses
the Davidson solver with a residual criterion of $10^{-8}$. These three
MoS$_2$ families use the dense reference eigensolver, for which the configured
Davidson tolerances are inactive. The screening and statistical criteria
are unchanged across these settings.

\section{Pseudopotentials}
\label{sec:pseudopotentials}

\subsection{HGH analytic parameters}
All Hartwigsen--Goedecker--Hutter (HGH) parameters were transcribed from
HGH98 Table I \cite{hartwigsen1998} and compared with the CP2K
\texttt{GTH\_POTENTIALS} data,
ABINIT test pseudopotentials and the Octopus HGH-LDA set.
Table~\ref{tab:hgh} lists the adopted values (radii in bohr, coefficients
in Hartree). Hydrogen uses the HGH98 parametrization distributed as
``GTH-LDA''/``GTH-PADE''; the original GTH96 hydrogen fit differs and is
documented but not adopted. Molybdenum uses the six-electron valence
partition (q6, Kr core, with the $4s/4p$ semicore states kept in the core);
the q14 semicore variant is not used, so the treatment of the $4s/4p$
states is part of the
operator definition.

\begin{table}[t]
\centering
\caption{Adopted HGH-LDA parameters (radii in bohr, coefficients in
Hartree). Values are retained at source precision because they define the
pseudopotential operator rather than measured outcomes. $h^l$ gives the
symmetric coefficient matrix upper triangle
$(h_{11}, h_{12}; h_{22})$ where two projectors exist; unlisted $C_i$
vanish.}
\label{tab:hgh}
\begin{tabular}{@{}llll@{}}
\toprule
Element & Channel & Radius & Coefficients \\
\midrule
H  & local & $0.2$        & $C_1 = -4.18023680$, $C_2 = 0.72507482$ \\
C  & local & $0.34883045$ & $C_1 = -8.51377110$, $C_2 = 1.22843203$ \\
   & $s$   & $0.30455321$ & $h^s_{11} = 9.52284179$ \\
Si & local & $0.44$       & $C_1 = -7.33610297$ \\
   & $s$   & $0.42273813$ & $(5.90692831, -1.26189397; 3.25819622)$ \\
   & $p$   & $0.48427842$ & $h^p_{11} = 2.72701346$ \\
S  & local & $0.42$       & $C_1 = -6.55449184$ \\
   & $s$   & $0.36175665$ & $(7.90530250, -1.73188130; 4.47169830)$ \\
   & $p$   & $0.40528502$ & $h^p_{11} = 3.86657900$ \\
Mo & local & $0.699$      & $C_1 = 7.99586821$ \\
   & $s$   & $0.67812595$ & $(1.28960728, -0.38656751; 0.99811301)$ \\
   & $p$   & $0.80077140$ & $(0.30141226, -0.31338933; 0.74161451)$ \\
   & $d$   & $0.45338393$ & $(-2.80970771, 3.00775463; -6.82094635)$ \\
Al & local & $0.45$       & $C_1 = -8.49135116$ \\
   & $s$   & $0.46010427$ & $(5.08833953, -1.03784325; 2.67969975)$ \\
   & $p$   & $0.53674439$ & $h^p_{11} = 2.19343827$ \\
\bottomrule
\end{tabular}
\end{table}

Unlisted $C_i$ coefficients vanish. Carbon ships no p projector: HGH98
Table I lists $r_p = 0.23267730$ bohr with $h^p_{11}$ exactly zero, the p
scattering being absorbed into the local part.

\paragraph{Off-diagonal coefficients (HGH98 Eq.~(25)).}
The off-diagonal $h^l_{12}$ values are not independent fit parameters:
HGH98 Eq.~(25) fixes them analytically from the diagonals
($h^s_{12} = -\tfrac12\sqrt{3/5}\, h^s_{22}$, with the $l = 1, 2$
analogues). The stored values reproduce the CP2K file to its 8-digit
transcription precision (e.g.\ silicon
$-\tfrac12\sqrt{3/5} \times 3.25819622 = -1.2618939699\ldots$ vs.\
$-1.26189397$). The full HGH98 recurrence, rather than a diagonal-only
approximation, is used throughout.

\subsection{PSP8 radial channels}
The parser follows the ABINIT PSP8 format specification and reader conventions
\cite{vansetten2018}. It uses one shared linear radial mesh with a
$10^{-9}$-bohr spacing tolerance and Bloechl--KB projectors with
$D = \mathrm{ekb}$ and no renormalization. The implementation supports
$l_{\max} \leq 3$, excludes spin--orbit extensions, stores but does not apply
the model-core-charge (NLCC) and pseudo-valence blocks, and ignores the ONCVPSP
\texttt{<INPUT>} trailer.

A PseudoDojo NC-SR v0.4 silicon PSP8 file in the LDA, with
$l_{\max}=2$ and six KB channels, was used in a large-box $\Gamma$-point
SCF calculation. The calculation converged in 39 iterations and satisfied
the band-sum identity to $2.7\times10^{-15}$. This file contains a nonzero
model-core charge and an extended pseudo-valence-charge block. These terms are
stored by the parser but omitted from this operator construction. The
calculation therefore tests the PSP8 reader, KB assembly, and band-sum
identity. The same-operator physical comparison in
Sec.~\ref{sec:physical-reference} uses the complete analytic HGH operator.

\section{Screening tests and response-shell concentration}
\label{sec:screen-tests}

\paragraph{Gapped-reference contrast and metal-case saturation.}
On the graphene Dirac path, a gapped reference with a Fermi gap
$43{,}000\times$ larger reaches a distance-to-bound ratio of $0.933$
against a value printed as $1.000$ at the Dirac contact; tightness saturates at
$0.88$--$0.98$ above the gap within the declared trial families. On the
converged fcc Al HGH spectrum the same check reproduces the $\tau/\Delta$
saturation band $0.97$--$0.98$ above the local pair gap; unlike graphene,
no $k$ point pins a pair at exact half occupation, and the tightest values
appear in the high-temperature half-occupation limit instead.

\paragraph{Global-k per-case quantiles.}
On the graphene full-path case with the shared chemical potential ($310$
trials, $\tau = 0.02$--$1.0$~Ha), the distance-ratio quantiles are
p50/p90/max $0.813/0.982/1.000$ to the displayed precision; the
double-precision saturation floor at the
low-temperature end is $1.59\times10^{-2}$~Ha. On the converged fcc Al
64-point $k$-mesh spectrum ($400$ trials), the shared chemical potential is
$0.290$~Ha, compared with the self-consistent value $0.285$~Ha. No
distance, gap, or trace-criterion violation occurs, and the p50/p90/max ratios
are $0.668/0.940/0.960$. The shared constraint therefore gives a looser
bound than the separate-window construction in these data. This Al kernel
test combines the 12 retained eigenvalues with seeded random orthogonal
frames; it is distinct from the in-loop Ritz compression defined below.

\paragraph{Per-case verification quantiles.}
The full quantile series behind the screening ECDF panels extend the medians
of Section~\ref{sec:numerical-verification}. For the randomized check
($10{,}000$ samples), the entries are rounded to three decimal places and
ordered as
$(q_{50},q_{90},q_{99},q_{\max})$:
\begin{align*}
Q_d&=(0.632,1.000,1.000,1.000),\\
Q_F&=(0.614,0.997,1.000,1.000).
\end{align*}
Distance and free-energy ratios are undefined for 114 and 434 ensemble
members, respectively, and are omitted from the corresponding summaries.
For the graphene Dirac-path case (164 trials) and the fcc Al 64-k-point metal
case (1,614 trials), the entries are ordered as
$(q_{50},q_{90},q_{\max})$:
\begin{align*}
Q_d^{\mathrm{graphene}}&=(0.791,0.979,1.000), &
Q_F^{\mathrm{graphene}}&=(0.743,0.975,0.999),\\
Q_d^{\mathrm{Al}}&=(0.779,0.998,1.000), &
Q_F^{\mathrm{Al}}&=(0.748,0.998,1.000).
\end{align*}
The global-k graphene ensemble contains two empty distance cells under the
same convention.

\paragraph{Probe-ensemble quantiles.}
\label{sec:probe-ensembles}
The adaptive-direction test and the reciprocal-shell calculation use
different probe ensembles on the physical HGH silicon operator.
In the former, three-mode bases leave more than $64\%$ of the response
energy outside the retained span; $90\%$ coverage requires $14$ of
the $16$ measured directions.

The exploratory reciprocal-shell calculation instead uses $128$ Fourier
derivative columns at a $6$~Ha cutoff, the $\Gamma$ point and a $13^3$ grid.
The fitted response-energy decay is approximately $1.529$ decades per
bohr$^{-1}$ of $|\mathbf G|$.
Its cumulative derivative-energy share reaches $0.927$ at
$|\mathbf G|=2.031$~bohr$^{-1}$ and $0.992$ at
$|\mathbf G|=3.000$~bohr$^{-1}$, exceeding $90\%$ and $99\%$, respectively.
In the corrected fcc silicon cell, the self-shell shares range from
$0.141$ to $0.590$ and vary nonmonotonically with shell index.

Section~\ref{sec:shell-concentration} uses $168$ derivative columns at
$25$~Ha, a $4^3$ $k$ mesh and a $33^3$ grid.
There, the cumulative shares are $0.982$ at $3.000$~bohr$^{-1}$ and
$0.993$ at $3.182$~bohr$^{-1}$, so the $99\%$ threshold occurs at the
latter shell.
Both calculations measure cumulative derivative-energy shares, but their
different discretizations and probe ensembles give different thresholds.
Their shell radii follow the same relation
$|\mathbf G|=(2\pi/a)\sqrt S$, with $a=10.260$~bohr and shell index $S$.

\subsection{Silicon response-shell concentration}
\label{sec:shell-concentration}

The fixed-parameter calculation reports $715$ shells over $168$ finite-difference
derivative columns with closure relative error $3.630\times10^{-15}$
(the columns are recomputed at the finest finite-difference level). The
shell truncation $S = 27$, in units of $(2\pi/a)^{2}$, selects
$168$ basis columns across ten shells: the $99\%$-coverage shell has
index $27$, and the next-smaller truncation $S = 24$ covers
$0.982$
(Table~\ref{tab:shell-sequence}); the $400$-column cap is not binding. The modes per
retained shell are $\{3{:}8, 4{:}6, 8{:}12, 11{:}24, 12{:}8, 16{:}6,
19{:}24, 20{:}24, 24{:}24, 27{:}32\}$, summing to $168$. The cumulative
derivative-energy share reaches $90\%$ within
$|\mathbf{G}|^{2} \le 11$ (cumulative $0.904$, four shells) and
$99\%$ within $|\mathbf{G}|^{2} \le 27$ (cumulative $0.993$, ten shells); the
finest-rung complement leakage of the baseline Jacobian projection, which is
distinct from the shell-coverage complement above, is
$8.55\times10^{-2}$. Table~\ref{tab:shell-sequence} reports the ten
retained shells. The saved output lists $142$ non-negligible shells;
the closure value above is the recorded full-shell accumulation, not a
sum reconstructed from those printed entries.

\paragraph{Strongest-coupling soft-mode composition.}
The strongest-coupling soft residual singular value is $0.854$, equal to the
strongest-coupling fixed-point margin in
Section~\ref{sec:intervention-results}. Its six largest shell weights are
$0.908$ for shell 3, $0.047$ for shell 4, $0.032$ for shell 8,
$0.004$ for shell 12, $0.004$ for shell 11, and $0.002$ for shell 19.
The softest direction is therefore dominated by the lowest retained shell.

The converged reference has $F=-7.942$~Ha,
$\mu=0.228$~Ha, 11 occupation-selected bands, first omitted
occupation $3.563\times10^{-11}$ and density residual
$9.129\times10^{-10}$. The associated intervention results are reported in
Section~\ref{sec:intervention-results}.

\begin{table}[tbp]
\centering
\caption{Derivative-energy shares of the ten retained reciprocal shells in
silicon. $|\mathbf{G}|^2$ is in units of $(2\pi/a)^2$; cumulative shares are
ordered by increasing shell index.}
\label{tab:shell-sequence}
\begin{tabular}{@{}rrrr@{}}
\toprule
$|\mathbf{G}|^2$ & $|\mathbf{G}|$ (bohr$^{-1}$) & share & cumulative \\
\midrule
$3$  & $1.061$ & $0.509$ & $0.509$ \\
$4$  & $1.225$ & $0.237$ & $0.745$ \\
$8$  & $1.732$ & $0.100$ & $0.846$ \\
$11$ & $2.031$ & $0.058$ & $0.904$ \\
$12$ & $2.121$ & $0.034$ & $0.938$ \\
$16$ & $2.450$ & $0.003$ & $0.941$ \\
$19$ & $2.669$ & $0.017$ & $0.959$ \\
$20$ & $2.739$ & $0.017$ & $0.976$ \\
$24$ & $3.000$ & $0.007$ & $0.982$ \\
$27$ & $3.182$ & $0.010$ & $0.993$ \\
\bottomrule
\end{tabular}
\end{table}

\section{Cross-lineage regression of recorded work counters}
\label{sec:work-prediction}

\begin{figure}[tbp]
\centering
\includegraphics[width=\textwidth]{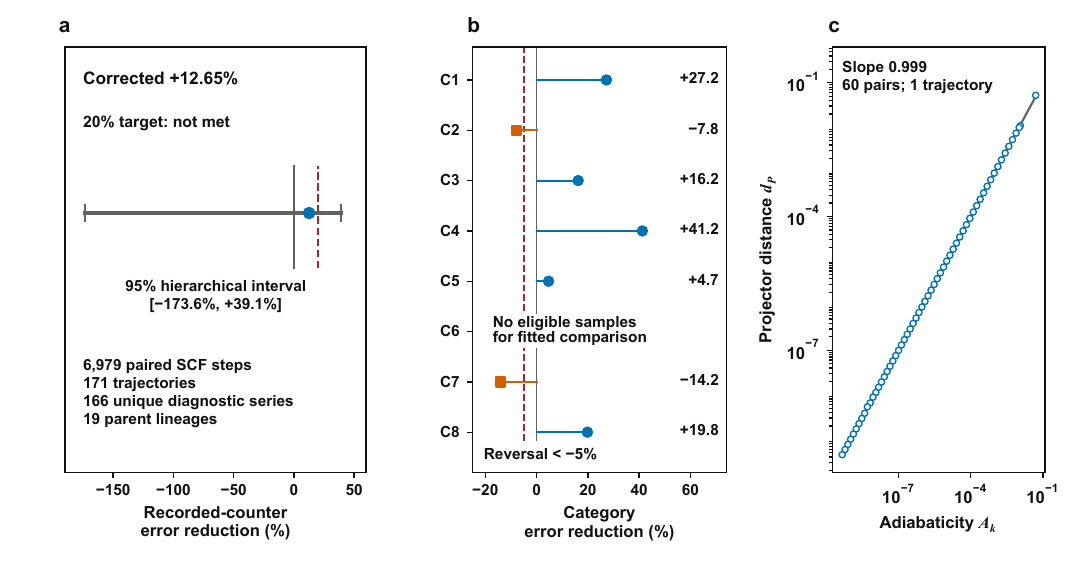}
\caption{Local projector geometry and cross-lineage counter regression.
\textbf{(a)} Intercept-corrected leave-one-parent-out relative RMSE
reduction for the original recorded-counter target against the best
observed baseline among six candidates:
$12.65\%$, with a hierarchical-bootstrap 95\% percentile interval
$[-173.6\%,39.1\%]$ from $10{,}000$ replicates.
The point estimate is below the prespecified $20\%$ target (dashed line);
the interval includes worsening and gains above that target.
The final fit uses $6{,}979$ paired SCF steps from $171$ trajectories,
corresponding to $166$ unique diagnostic series within $19$ parent lineages.
The bootstrap samples category, parent, diagnostic series and individual
step indices hierarchically, refitting all models and reselecting the
baseline in every replicate. Duplicate series share a sampling identity
but retain both copies of their rows. The percentile interval describes
this resampling procedure; it does not establish coverage for arbitrary
serial dependence or for a homogeneous per-step work target.
\textbf{(b)} Corrected category-level reductions. C1--C8 follow
Table~\ref{tab:work-category}; squares mark pseudopotential hydrides (C2)
and polar insulators (C7), below the $-5\%$ reversal threshold.
Segments connect estimates to zero; they are not confidence intervals.
The nonmagnetic-metal category (C6) has no samples eligible for this fitted
comparison, not a fitted zero effect.
\textbf{(c)} Projector distance $d_P$ versus $A_k$ for all 60 positive pairs
on one converged trajectory. The descriptive log--log fit has slope $0.999$.
This single-trajectory check is distinct from the grouped prediction test.
AI-assisted figure preparation was checked against the numerical records
and mathematical definitions.}
\label{fig:si-work-prediction}
\end{figure}

This auxiliary experiment tests a zero-temperature, gapped-subspace
diagnostic, $A_k$, built from integer occupations and occupied--unoccupied
energy denominators. Here $k$ labels the SCF step, not a Brillouin-zone
point. The diagnostic measures first-order occupied-projector motion under
a potential change; it is neither the finite-temperature canonical
derivative nor the fixed-$H$ tangent residual $r(D)$ used for screening.
The regression and the screening bound address different quantities.
Figure~\ref{fig:si-work-prediction} reports the grouped result and a
separate local scaling check.

The recorded work target requires a distinction between kernels. The
pseudopotential counter accumulates Davidson matrix--vector columns across
iterations, whereas the reduced-model counter measures one subspace-reuse
attempt. The latter excludes the accompanying dense reference solve.
Their pooled logarithms therefore do not define a common per-step cost.
We retain this original response variable to reproduce the reported
regression, rather than reinterpret it as total eigensolver work.
Predictors and responses also come from the same completed SCF record;
the full feature vector is not a pre-solve forecast.

\subsection{Design and local calibration}
The grouped calculation contains eight physical test categories, three independent
parent lineages per category and ten descendants per parent. Each parent is
crossed with five strain values, $\{-4\%,-2\%,0,+2\%,+4\%\}$, and two mixing
settings, giving 240 trajectories. A parent is defined by the electronic
kernel, composition, structure prototype and initial-condition family; strain,
mixer and individual SCF steps are descendant-level variations. The primary
evaluation leaves out all rows of one parent at a time.

The five baseline covariates are $-\ln\gamma_k$, $\ln v_k$,
$\ln q_k$, $\ln(1+k)$ and $\ln N_{\mathrm{at}}$, where $\gamma_k$
is the anchor gap, $v_k$ is the potential-change infinity norm,
$q_k$ is the recorded density-residual scalar and $N_{\mathrm{at}}$
is the atom count. Numerical gap and potential values are in hartree.
Five single-covariate fits and their joint fit define the six baselines.
The main model adds $\ln A_k$ to the joint baseline. The ablation further
adds $|\ln A_k-\ln A_{k-1}|$, using consecutive valid records before
the inclusion filter. A separate geometry-only fit uses $\ln A_k$.
Every corrected fit includes an intercept and minimizes
\begin{equation}
 \sum_{i\in\mathcal T} w_i(y_i-\mathbf{x}_i^{\mathsf T}\boldsymbol\beta)^2
 +10^{-10}\|\boldsymbol\beta\|_2^2,
\end{equation}
with the same small diagonal term applied to the intercept.
Here $\mathcal T$ excludes the held-out parent or category, and $w_i$
is one for point estimates or the sampled row multiplicity for resampling.
The two responses are the natural logarithms of the recorded work counter
and projector distance. Baseline selection minimizes the pooled held-out
RMSE, rather than using a nested training-only selection procedure.

\paragraph{Local projector relation and subspace reuse.}
\label{sec:projector-reuse}
On one independent converged H$_2$ trajectory, all 60 positive residual pairs enter
the log--log regression of the normalized occupied-projector distance $d_P$ on
coupling-weighted adiabaticity $A_k$. The slope is $0.999$, the mean
$d_P/A_k$ ratio is $1.000$ to the displayed precision, and the mean relative
deviation is $0.005$.
Of the 60 reuse attempts, 54 satisfy the retained-state spectral-separation
condition obtained from Weyl's eigenvalue perturbation inequality,
\begin{equation}
 \lambda_{m+1}(H_c)-\lambda_m(H_c)-2\|v_k-v_c\|_\infty>0,
 \label{eq:retained-weyl-separation}
\end{equation}
where $H_c$ and $v_c$ define the reference state and $m$ is the retained
subspace dimension.
Only 29 attempts pass the full provisional reuse test, which also checks
candidate-residual convergence, candidate separation, a density-error
bound and orthogonality.
The other 25 spectrally separated attempts fail candidate-residual convergence.
Thus spectral separation alone does not imply acceptance, and the 29
passes do not count tests of the local $d_P/A_k$ relation.
Acceptance occurs late in this single trajectory and is confounded with
iteration position.

The first-order relation departs by 10\% at $A_k\simeq0.30$ in the reduced
model. Step halving in eight directions gives slopes in
$[0.999,1.001]$. The same deviation criterion gives category-specific
$A^\ast$ values of $0.023$ (pseudopotential hydrides), $0.031$ (covalent
semiconductors, a lower bound because the crossing is not reached), $0.025$
(semimetals), $0.072$ (two-dimensional semiconductors), $0.165$ (polar
insulators) and $0.096$ (molecular solids), with inclusion defined by
$A_k\le0.85A^\ast$. The nonmagnetic-metal category
is excluded from this calibration because its gap closes; it contributes
to the coverage summary but not the fitted work comparison.
The projector distance is evaluated
in the same retained unoccupied subspace as $A_k$. This removes a nonnegative
truncation tail of up to 85\% and changes the median $d_P/A_k$ deviation from
$0.284$ to $1.0\times10^{-4}$ without changing the SCF trajectories.

\subsection{Denominators and grouped results}
Of 240 attempts, 216 converge. Of the remaining 24, 23 are recorded as
nonconverged and one ends without a convergence result or a saved trajectory.
These 24 attempts occur in
toy diatomics (3), semimetals (5), two-dimensional semiconductors (5),
nonmagnetic metals (5) and polar insulators (6); 21 of the 24 belong to the linear
mixing-$0.70$ arm. The 239 available trajectories contain
$15{,}610$ recorded SCF steps: $12{,}160$ from converged trajectories and
$3{,}450$ from the other available trajectories. The missing trajectory
precludes an exact step count over all attempts. Attempts without recorded
convergence are filtered out
before step-level diagnostics and coverage are assembled. Of the converged
steps, $7{,}968$ enter the diagnostic-eligibility denominator,
$7{,}039$ lie inside the calibrated $A^\ast$ domain, and $6{,}979$ paired
rows are used by every model. Thus $0.883=7{,}039/7{,}968$ is a
conditional inclusion fraction, not coverage of all attempted steps.
Nineteen pairs of converged trajectories are
numerically identical because both mixer arms use the same parameter for those
parents. Assigning each pair a common sampling identity leaves 197 distinct converged
streams; the final included set contains 171 trajectories and 166 distinct
streams nested within 19 parent lineages. Point estimates use the same
6,979 paired rows. The initial design has 24 parents; merely removing the
metal category would leave 21, but only 19 contribute to the final fit.
Missing gapped diagnostics and nonconvergence may depend on problem
difficulty, so these conditional results do not estimate performance over
all attempted trajectories.

\begin{table}[tbp]
\centering
\caption{Category-level relative error reductions and inclusion fractions for
the original cross-lineage counter regression. Inclusion is conditional on the
diagnostic-eligible converged-step denominator. A negative value denotes
worse prediction than the strongest baseline. The metal category has no
fitted effect because no step is admitted by the gapped-diagnostic rule.}
\label{tab:work-category}
\begin{tabular}{@{}lrrr@{}}
\toprule
category & prespecified fit (\%) & intercept-corrected fit (\%) & inclusion \\
\midrule
toy diatomics & $+18.9$ & $+27.2$ & $0.997$ \\
pseudopotential hydrides & $-11.6$ & $-7.8$  & $0.911$ \\
covalent semiconductors & $+2.1$  & $+16.2$ & $0.937$ \\
semimetals & $+19.4$ & $+41.2$ & $0.924$ \\
two-dimensional semiconductors & $+5.9$  & $+4.7$  & $0.941$ \\
nonmagnetic metals & -- & -- & $0.000$ \\
polar insulators & $-12.6$ & $-14.2$ & $0.952$ \\
molecular solids & $-5.7$ & $+19.8$ & $0.977$ \\
\bottomrule
\end{tabular}
\end{table}

The prespecified through-origin fit reduces grouped held-out error by 8.40\%,
below the required 20\%, with a hierarchical-bootstrap 95\% interval of
$[-78.1\%,+30.0\%]$ from $10{,}000$ parent-level replicates. Three category
effects are below the $-5\%$ reversal threshold. A subsequent correction adds
the omitted intercept and merges the 19 identical trajectory pairs during
resampling; it retains both copies of their rows and leaves the inclusion
domain unchanged. The
corrected reduction is 12.65\%, with interval
$[-173.6\%,+39.1\%]$, and the reversals narrow to pseudopotential hydrides
and polar insulators
(Table~\ref{tab:work-category}). Trajectory-, parent- and category-equal
scoring gives 8.7\%, 11.1\% and 9.5\%, respectively, all below 20\%.
These summaries reweight the same held-out errors, without refitting or
reselecting the pooled baseline.
The corrected main and ablation models have similar RMSEs
($2.055$ and $2.053$; paired difference interval
$[-0.066,+0.131]$), while the strongest baseline has RMSE $2.353$.
Independent reconstruction of all $10{,}000$ draws reproduces the displayed
main-improvement and paired-difference intervals with the original
row-ordered accumulation and elimination procedure.
Changing the accumulation and linear-solution order changes 1,023 main-model gains by more than
$10^{-6}$ and moves the lower percentile from $-173.645\%$ to $-170.223\%$.
SVD checks on 14 selected draws, chosen by arithmetic disagreement or
proximity to the lower percentile, identify 16 rank-deficient training
matrices among 230 fold fits of the main and ablation models. These selected checks are not
a complete SVD-based uncertainty calculation.
A normal--normal sensitivity calculation estimates zero between-category
variance ($Q=1.19$ on 6 degrees of freedom). With at most three contributing
parents per category, the data do not resolve whether the reversals are
physical or statistical.

The leave-one-category-out calculation is also negative for the recorded
counter: relative improvement $-0.224$ after the intercept correction
($-0.544$ in the prespecified fit), compared with $+0.990$ for the local
projector-distance target. The local perturbative relation therefore remains
accurate within its calibrated domain. The pooled counter regression misses
the prespecified improvement criterion, but its mixed response definitions
do not test a universal law for per-step eigensolver work.

\subsection{Incremental-counter sensitivity}
We separately refit the pseudopotential subset to distinguish cumulative
counts from increments. All $5{,}813$ selected rows have consecutive raw
predecessors and strictly positive differences $W_k-W_{k-1}$.
Differences are computed before diagnostic filtering; an omitted row is
never treated as an omitted increment. This subset contains 144 trajectories,
139 distinct series and 16 parents across six categories.
The new target is $\ln(W_k-W_{k-1})$, with the same predictors,
inclusion domain and held-out groups. It measures counted iterative
applications, excluding dense solves and other uncounted operations.
The interpretation follows the documented accumulation rule and the saved
counter sequences, rather than an independent recount of Hamiltonian applications.

\begin{table}[tbp]
\centering
\caption{Post-hoc endpoint and duplicate-weight sensitivity within the
pseudopotential subset. Both columns report $100(1-\mathrm{RMSE}_{\mathrm{main}}/
\mathrm{RMSE}_{\mathrm{baseline}})$; negative values denote larger errors.
Baselines are selected separately for each response and grouping.
The last row removes repeated copies of identical series before fitting.
These are descriptive refits of existing records, not new independent trials.}
\label{tab:counter-sensitivity}
\begin{tabular*}{\textwidth}{@{\extracolsep{\fill}}lrrr@{}}
\toprule
response and weighting & rows & parent holdout (\%) & category holdout (\%) \\
\midrule
recorded cumulative count & 5,813 & $+4.429$ & $-426.133$ \\
increment, original rows & 5,813 & $+3.638$ & $-454.743$ \\
increment, unique series & 5,143 & $+3.429$ & $-356.199$ \\
\bottomrule
\end{tabular*}
\end{table}

For the incremental target, the parent-held-out RMSEs are $1.172$ for
the main model and $1.216$ for the joint baseline.
Leaving out complete categories gives $7.705$ and $1.389$, respectively,
with the residual-only baseline selected in this comparison.
Removing 670 repeated rows changes these values to $1.239$ versus $1.283$
and $6.837$ versus $1.499$.
The modest parent-held-out gains therefore do not extend to held-out
material categories. They also remain retrospective associations, because
the predictors include information from the completed iteration.
Augmented-design SVD and normal-equation solutions agree in all displayed
RMSE digits; the large category-held-out errors are not removed by changing
the linear solver.

\section{Physical reference comparisons}
\label{sec:physical-reference}

\paragraph{Same-operator equation of state.}
Under a third-order Birch--Murnaghan fit, PySCF \cite{sun2020pyscf} gives
$a_0=5.403$~\AA{}, $B_0=93.7$~GPa, and $B'=4.21$ at zero smearing, and
$a_0=5.407$~\AA{} and $B_0=89.1$~GPa for the finite-temperature protocol.
At the same finite temperature, IDMate gives $a_0=5.384$~\AA{},
$B_0=91.5$~GPa, and $B'=4.454$. The same-operator differences are
$-0.023$~\AA{} ($-0.417\%$) and $+2.36$~GPa ($+2.65\%$). The separate
VASP-PAW fit gives $a_0=5.415$~\AA{} and $B_0=96.5$~GPa; relative to this
cross-operator reference, the IDMate differences are $-0.031$~\AA{}
($-0.566\%$) and $-5.01$~GPa ($-5.19\%$). Absolute total-energy zeros are
excluded from both comparisons.

\paragraph{Same-temperature silicon gap comparison.}
For the $4^3$ mesh at $\tau=0.02$~Ha, the recorded gaps are
$41.640$~mHa for IDMate at 25~Ha cutoff and $43.320$~mHa for PySCF.
The absolute difference from the unrounded values is $1.680$~mHa
($45.711$~meV). The PySCF zero-temperature value, $42.159$~mHa,
does not provide a matched-temperature comparison. Temperature matching
for the separate $2^3$ and $6^3$ comparisons has not been established,
so they are not used as matched validation here.
At fixed $4^3$ mesh and $\tau=0.02$~Ha, increasing the IDMate cutoff
through 10, 15, 20 and 25~Ha gives gaps of
$41.949$, $41.758$, $41.684$ and $41.640$~mHa, respectively.
The sequence moves away from, not toward, either PySCF value.
Matching the temperature alone does not isolate the remaining discrepancy:
basis and integration-grid convergence must also be checked.

\paragraph{Selected-pair comparison detail.}
At the (25~Ha, $4\times4\times4$) point selected by the $k$-mesh rule,
Birch--Murnaghan $B'$ is $4.454$ for IDMate and $4.241$ for the VASP-PAW
refit. The $\Delta$ values are $6.97$~meV/atom against VASP-PAW and
$4.82$~meV/atom against the finite-temperature, same-operator PySCF anchor.
The equation-of-state shape maximum is $14.02$~meV/atom over $41$ samples.
The own-reference-volume comparison gives a near-Fermi maximum error of
$0.250$~eV over five band pairs and $155$ points. The IDMate and VASP-LDA path
gaps are $0.486$ and $0.439$~eV
for the indirect gap and $2.582$ and $2.503$~eV for the minimum direct gap.

\paragraph{Same-volume band comparison.}
The main-text band metric compares both implementations at the experimental
lattice constant, $a = 5.431$~\AA{}, retaining the respective numerical
settings of the two implementations. The mean absolute error is
$0.034$~eV and the maximum is $0.103$~eV over the same five band pairs. Every
occupied pair agrees within $0.015$~eV, while the residual is concentrated in
the conduction pairs (VBM$+1$: $0.051$~eV; VBM$+2$: $0.076$~eV). At this
volume the IDMate
indirect and minimum direct gaps stand $+0.090$ and $+0.066$~eV above the
VASP values ($0.529$/$2.569$ vs $0.439$/$2.503$~eV). The earlier
own-referenced-volumes comparison of the preceding paragraph has a mean
absolute error of $0.060$~eV, maximum $0.250$~eV, and gap differences
$+0.047$/$+0.080$~eV. The indirect-gap change between these comparisons
is $+43$~meV. These comparisons do not isolate an HGH-versus-PAW operator
contribution. The VASP inputs use Gaussian smearing
(\texttt{ISMEAR=0}, width $0.05$~eV) and a $\Gamma$-centered $4^3$ mesh;
IDMate uses Fermi--Dirac occupations at $\tau=0.02$~Ha
($0.544$~eV) and the half-shifted even Monkhorst--Pack mesh.
Smearing, quadrature and operator differences remain confounded.
Moreover, the 20-to-25~Ha cutoff change in the silicon energy is about
$2.46$~meV/atom, above the stated $1$~meV/atom convergence target.
The VASP results are therefore shape-level comparisons, not a demonstration
of operator equivalence or uniformly converged VASP-level accuracy.

\paragraph{Reference calculations for the three anchor systems.}
The reference calculations use the Ceperley--Alder/Perdew--Zunger
parametrization of the LDA \cite{ceperley1980,perdew1981}, matching the
exchange-correlation family and parametrization of the verified functional
switch. For silicon, a third-order Birch--Murnaghan refit of the five-volume
VASP 6.1.0 PAW calculation (ENCUT 400~eV, $4\times4\times4$) has an RMS
residual of $0.18$~meV. It gives an equilibrium lattice constant of
$5.415$~\AA{} (experiment $5.431$), a bulk modulus of $96.5$~GPa
(experiment ${\sim}98$), and $B' = 4.24$. A parabolic $E(V)$ fit instead
gives $5.458$~\AA{} and $97.0$~GPa but misfits the five points at
$26$~meV RMS. For
graphene the reference places the Dirac contact at K exactly at the Fermi
level; the bands at the stated cutoff reproduce the Dirac cone with a Dirac-window
mean absolute deviation of $0.174$~eV and cone slopes steeper by
$5$--$7\%$. For MoS$_2$ the K-point direct gap at the stated cutoff
is $1.851$~eV against the reference $1.778$~eV, and the
valence-band-maximum-aligned comparison gives valence and conduction mean absolute deviations of
$0.053$~eV and $0.058$~eV at this cutoff. The graphene and MoS$_2$
comparisons are shape-level tests and are not cutoff-converged reference
calculations.

\section{Cross-family transfer and work tests}
\label{sec:cross-family-work-tests}

The H$_2$/He$_2$/LiH soft-Coulomb families were used to test whether a
single data-driven policy improves the worst family. A mixed-family trajectory
forest selected only He$_2$ cases; H$_2$ and LiH abstained, leaving the
worst-family iteration ratio at $1.0$. A no-slowdown tree likewise selected no
LiH cases and gave held-out ratios of $1.212$ for H$_2$ and $1.007$ for
He$_2$, again leaving the worst-family value at $1.0$.

Abstention is not the sole cause. On the LiH single-bond test, the
secant--Kerker variant takes 77 iterations, compared with 44 for the linear
reference, or $1.75$ times as many iterations. The measured wall-time ratio is
$t_{\mathrm{ref}}/t_{\mathrm{secant}}=0.578$ (40.226~s/69.631~s).
The secant-preconditioner
schedule gives a LiH family ratio of $0.713$ and one LiH case does not
converge within 320 iterations. In the 60-case online calculation, the
per-family ratios are $1.131$ (H$_2$), $0.997$ (He$_2$) and $1.0$ (LiH).
One He$_2$ case increases from 16 to 25 iterations; five timing repetitions
increase from a mean of $3.68$ to $5.78$ ms.

A grouped value-of-information calculation has nine of the ten required
supergroups at its 90\% inclusion target and therefore yields no estimate at
that target. At an 80\% diagnostic threshold it selects no additional
calculations in either out-of-fold or transfer evaluation, giving zero net
gain. No tested policy in these auxiliary calculations improves the
worst family beyond $1.0$.

\section{Rank-one intervention and mixer response}
\label{sec:intervention-results}

\subsection{Reduced response models}
For the H$_2$/Grid4 model, $\lambda_c=-59.076$~Ha\,bohr$^3$. At coupling strengths
$0.1\lambda_c$, $0.2\lambda_c$, and $0.4\lambda_c$, the measured Jacobian changes differ from
the rank-one prediction by $1.47\times10^{-6}$,
$1.47\times10^{-6}$, and $1.46\times10^{-6}$, respectively. The
intrinsic fixed-point margin moves from $0.876$ to $0.797$ and
$0.568$. Independent reconvergence gives
$\|\Delta\rho\|_W=1.79\times10^{-10}$~bohr$^{-3/2}$,
$\|\Delta V\|_W=1.37\times10^{-9}$~Ha\,bohr$^{3/2}$, and
$|\Delta F|=5.07\times10^{-11}$~Ha.

In the three-mode $\Gamma$-point plane-wave model, the finite-difference
Jacobian has maximum charge leakage $3.61\times10^{-13}$. The rank-one
errors are $7.72\times10^{-6}$ and $1.02\times10^{-5}$ at $0.2\lambda_c$ and
$0.4\lambda_c$, with central-difference order $2.00$. Independently
reconverged susceptibilities have orders $1.999$--$2.000$ and differ from
the projected linear prediction by $1.46\times10^{-5}$--$2.29\times10^{-5}$;
the selected response grows by a factor of $1.67$. The projected singularity
is bracketed by residual determinants $+0.138$ and $-0.137$. The
Hellmann--Feynman derivative and second-energy-curvature errors are
$2.77\times10^{-9}$ and $1.26\times10^{-8}$.

The quoted critical couplings belong to distinct models and are not directly
comparable: $-59.076$~Ha\,bohr$^3$ for H$_2$/Grid4,
$-47.918$~Ha\,bohr$^3$ for the
three-mode $\Gamma$-point model, and the mesh-dependent values in
Table~\ref{tab:k2-nested-rungs} for the weighted-$k$ model.

\begin{table}[tbp]
\centering
\caption{H$_2$/Grid4 spectral radii $\rho(M_B)$ for four fixed
preconditioners and four intervention strengths.}
\label{tab:h2-mixer-radii}
\begin{tabular}{lcccc}
\toprule
operator & fixed $J$ & $0.1\lambda_c$ & $0.2\lambda_c$ & $0.4\lambda_c$ \\
\midrule
unit                     & $0.447$              & $0.447$  & $0.447$ & $0.447$ \\
scalar-0.25              & $0.772$              & $0.773$  & $0.793$ & $0.842$ \\
scalar-0.70              & $0.361$              & $0.364$  & $0.420$ & $0.557$ \\
exact linearized control & $1.11\times10^{-15}$ & $0.100$ & $0.200$ & $0.400$ \\
\bottomrule
\end{tabular}
\end{table}

The unit-preconditioned spectral radius is unchanged to the displayed
precision in this model. The two scalar
preconditioners move toward unit spectral radius as the intervention softens
the response. The exact linearized control follows the imposed coupling
fraction from a baseline radius of $1.1\times10^{-15}$. The corresponding
$g_{20}$ gains change from $0.577$, $0.786$, $0.438$, and
$1.11\times10^{-15}$ at zero coupling to $0.577$, $0.867$,
$0.648$, and $0.564$ at $0.4\lambda_c$.

\subsection{Silicon material calculation}
For HGH silicon at 25~Ha and a $4\times4\times4$ mesh,
$\lambda_c=-22.127$~Ha\,bohr$^3$ (re-estimated in the retained basis as
$-22.131$~Ha\,bohr$^3$). The measured rank-one update differs from the prediction by
$5.47\times10^{-7}$ and $4.42\times10^{-7}$ at $0.2\lambda_c$ and
$0.4\lambda_c$, with finite-difference order $2.105$. The update norms
are $0.388$ and $0.775$, and the fixed-point margin moves from
$0.944$ to $0.854$. Charge leakage is
$2.60\times10^{-11}$, $2.22\times10^{-11}$, and
$2.95\times10^{-11}$ at $0$, $0.2$, and $0.4\lambda_c$. Residual
determinants $+0.575$ at $0.9\lambda_c$ and $-0.573$ at
$1.1\lambda_c$ bracket the projected singularity. The strongest
susceptibility growth is $1.666$. Zero-field reconvergence requires 140
iterations and reaches residual $1.90\times10^{-10}$; the
Hellmann--Feynman derivative and curvature-identity errors are
$5.92\times10^{-9}$ and $1.40\times10^{-8}$.

The anchor density minimum, $2.77\times10^{-4}$, limits the largest
admissible finite-difference step to
$2.89\times10^{-3}$~Ha\,bohr$^{3/2}$. The
calculation therefore uses
\begin{equation*}
(\delta_1,\delta_2,\delta_3)
=(2.0\times10^{-3},1.0\times10^{-3},5.0\times10^{-4})\
\mathrm{Ha}\,\mathrm{bohr}^{3/2}
\end{equation*}
rather than a $4\times10^{-3}$~Ha\,bohr$^{3/2}$ step.

\begin{table}[tbp]
\centering
\caption{Silicon fixed-$B$ spectral radii $\rho(M_B)$ at each coupling
level ($\lambda_c=-22.127$~Ha\,bohr$^3$).}
\label{tab:silicon-fixed-b}
\begin{tabular}{lcccc}
\toprule
level & unit & scalar-0.25 & scalar-0.70 & inverse control \\
\midrule
$0$               & $1.033$  & $0.762$ & $0.423$ & $2.97\times10^{-15}$ \\
$0.2\lambda_c$   & $1.007$  & $0.762$ & $0.405$ & $0.200$ \\
$0.4\lambda_c$   & $0.998$ & $0.780$ & $0.399$ & $0.400$ \\
\bottomrule
\end{tabular}
\end{table}

The $g_{20}$ gains, in the same level order, are
\begin{align*}
g_{20}^{\mathrm{unit}}&=(1.995,1.211,1.056),\\
g_{20}^{0.25}&=(0.765,0.766,0.789),\\
g_{20}^{0.70}&=(0.516,0.502,0.510),\\
g_{20}^{\mathrm{inv}}&=(3.07\times10^{-15},0.225,0.450).
\end{align*}

\subsection{Discretization sensitivity}
In the $3\times3$ cutoff--grid factorial, the finest cutoff edge changes the
projected Jacobian by $2.55\times10^{-13}$ and the common-grid density by
$4.91\times10^{-16}$. The finest grid edge changes them by
$6.81\times10^{-8}$ and $9.97\times10^{-15}$. Both soft-subspace overlaps
are 1.0, and the strongest-coupling complement leakage at the finest point is
$2.79\%$. In the nested-mesh calculation, baseline-to-strongest margins are
$0.963\rightarrow0.792$, $0.987\rightarrow0.833$, and
$0.990\rightarrow0.828$. Rank-one errors are at most
$6.3\times10^{-6}$, susceptibility orders are $1.996$--$2.002$, and
projected-prediction errors are at most $1.19\times10^{-5}$. On the
$2^3\rightarrow4^3$ edge, the Jacobian and susceptibility change by
$1.832\times10^{-2}$ and $1.464\times10^{-2}$, while the soft-subspace
overlap is $1.000$ to the displayed precision.

An earlier $1.05$~Ha/$12^3$ calculation gave a spurious finite-difference
order of $-4.878$ because occupied eigenpair residuals reached
$5.04\times10^{-6}$ near a degeneracy. The safeguarded self-adjoint solver,
with $10^{-11}$ residual and orthonormality criteria, restores the expected
order $2.002$ without changing the physical inputs. Density and free-energy
changes of $1.491\times10^{-3}$ and $1.385\times10^{-3}$~Ha/electron on
the coarsest nested-mesh comparison measure Brillouin-zone quadrature error
and are not used as response-mechanism criteria.

\section{Standalone controller and screening tests}
\label{sec:controller-tests}

\subsection{Three-system controller calculation}
At each $k$ point, the controller counts the bands inside
$|\varepsilon-\mu|\leq18\tau$ and sets a common window dimension $m$ to
the largest count, capped by the smallest available band count across
$k$ points. It then selects the $m$ bands nearest $\mu$ at each point.
Consequently, the selected states need not all lie within $\pm18\tau$;
their actual spectral extent and minimum occupation margin determine the
numerical conditioning. An empty initial window or $m<2$ causes abstention.
The compressed Hamiltonian is
\begin{equation}
H_k^{W}=\operatorname{Herm}\!\left[(U_k^{W})^\dagger H_kU_k^{W}\right].
\label{eq:window-ritz}
\end{equation}
The first excluded band below the window must be full and the first above
it empty within the per-spin tolerance $10^{-6}$. Fully occupied or empty
states outside the window are permitted by these frozen-edge checks;
their saturation does not automatically trigger fallback. Saturation of a
retained occupation invalidates the logarithmic residual and causes
abstention. The embedded window target is
$N_W=N_e-2\sum_k w_kn_{k,<}$, where $n_{k,<}$ is the number of frozen
bands below the window. The reported certificate distance is
$[\sum_k w_k\|D_k-D_k^\star\|_F^2]^{1/2}$ for the real embedding of these
common-dimension Ritz blocks. The reference $D_k^\star$ is the canonical
solution of the compressed Hamiltonians at the window target, not generally
the projection of the full-space canonical state. The distance excludes
the complement and its coupling to the window
(Sec.~\ref{sec:window-counterexamples}).

The low-cutoff candidate performs one eigensolve at reduced plane-wave
cutoff using the current effective potential, zero-pads its orbitals into
the full basis, and recomputes a shared Fermi chemical potential. It is
not a converged low-cutoff SCF calculation. The subspace candidate instead
retains orbitals from a previously completed all-reference SCF solve of
the target system and performs a Ritz update with the current Hamiltonian,
again recomputing common occupations. It is an oracle control for testing
the screen, not an online candidate source for an unknown target.
Its reference preparation is not included as a deployment expense in the
historical cost totals below. The tables call these candidates
``low-cutoff'' and ``oracle subspace'', respectively.

Table~\ref{tab:three-system-controller} summarizes the silicon, graphene and
aluminum calculations. The coarse-$k$ proposal is rejected in every system.
The graphene low-cutoff proposal has an empty screening window and therefore
abstains. All other reported proposals are accepted. Final free energies agree
with the corresponding all-reference calculations by at least three orders of
magnitude more tightly than the $10^{-7}$~Ha comparison tolerance.

\begin{table}[tbp]
\centering
\caption{Proposal-level quantities in the three-system controller
calculation. Accepted entries report ranges when two proposals are available.
The coarse-$k$ rows give the bound and directly evaluated window distance of
the rejected proposal. $d_{\mathrm{full}}^{\max}$ is the largest full
matrix distance among accepted proposals.}
\label{tab:three-system-controller}
\small
\setlength{\tabcolsep}{5pt}
\begin{tabular}{@{}lccc@{}}
\toprule
quantity & Si & graphene & Al \\
\midrule
accepted bound & $[5.9\times10^{-10},3.1\times10^{-9}]$ &
$1.46\times10^{-4}$ & $[4.12,4.86]\times10^{-10}$ \\
accepted window distance & $[1.3\times10^{-10},2.9\times10^{-9}]$ &
$4.74\times10^{-11}$ & $[1.44\times10^{-9},2.17\times10^{-8}]$ \\
coarse-$k$ bound & $6.26$ & $18.2$ & $4.58$ \\
coarse-$k$ window distance & $0.329$ & $0.746$ & $0.056$ \\
$|\Delta F|$ (Ha) & $8.88\times10^{-16}$ & $8.35\times10^{-14}$ &
$1.60\times10^{-14}$ \\
$d_{\mathrm{full}}^{\max}$ & $0.774$ & $0.169$ & $0.330$ \\
\bottomrule
\end{tabular}
\end{table}

The absolute Frobenius tolerance is 0.05. The empty graphene low-cutoff
window has full density-matrix distance $2.13$. These full distances are not bounded
by the window theorem; they are measured by a separate reference-map
evaluation.

\subsection{Empirical numerical allowance}
\label{sec:historical-allowance}
The historical controller calculations use the uncorrected residual bound
$B_0=r_E/(4\tau)$ in the weighted real-embedding metric.
When a finite verification-side window distance is available, their
numerical comparison uses $B_0$ plus
\begin{equation}
s_{\mathrm{fp}} = 10^{-11}+\delta_{\mathrm{tr}}
+\frac{F_{\mathrm{fp}}}{4\tau},\qquad
F_{\mathrm{fp}}=2048\,\varepsilon_{\mathrm{mach}}\,\tau
\frac{\sqrt{2n_W}}{\eta},
\label{eq:screen-slack}
\end{equation}
where $n_W=m$ is the common number of retained window bands per $k$ point,
$\delta_{\mathrm{tr}}$ is the absolute global embedded-trace deviation,
and $\eta$ is the smallest distance of any retained per-spin occupation
from 0 or 1. This is an empirical numerical allowance, not an interval
enclosure or a universal weighted-trace correction. In particular,
$\delta_{\mathrm{tr}}$ alone need not bound the displacement of the
canonical reference in the weighted Frobenius metric.
For $\tau=0.02$~Ha, $n_W=11$ and an assumed edge occupation of order
$e^{-18}$, the floor contribution is $3.50\times10^{-5}$ before the
trace term. This illustrative value is not a bound for every selected
window: common-$m$ selection can reduce $\eta$ below this assumption.

For an explicit weighted-trace example, take weights $(0.01,0.99)$,
two spectra $(0,0)$ and $(-0.6,0.6)$~Ha, and $\tau=0.05$~Ha.
The target canonical state has $\mu=0$; a candidate canonical at
$\mu=4\times10^{-6}$~Ha has zero projected residual. Its embedded trace
deviation is $8.019\times10^{-7}$ and its embedded weighted distance
from the target is $4.000\times10^{-6}$. Even enlarging the floor estimate
from two bands per block to $n_W=4$ gives only $8.543\times10^{-7}$ for
Eq.~\eqref{eq:screen-slack}, below the true distance. This disproves a
universal interpretation of the allowance, not the exact-trace theorem.
It also does not demonstrate false acceptance by the full controller,
whose separate charge checks may reject such a candidate.

Direct evaluation of this example in the screening routine gives
$r/(4\tau)=4.683\times10^{-12}$ and an actual allowance of
$8.390\times10^{-7}$ for the selected two-band blocks.
The measured window distance is $4.000\times10^{-6}$, so the subsequent
bound-verification assertion fails even though the distance is below
the operational $0.05$ threshold. A separate charge-guard test using this
candidate's electron number rejects the charge drift; it does not reconstruct
the full real-space proposal.

The common-dimension selection also requires the actual occupation margin.
For weights $(0.5,0.5)$, $\mu=0$, $\tau=0.01$~Ha and spectra
$(-0.3,0,0.3)$ and $(-0.01,0,0.01)$~Ha, the nominal counts are one and three.
The routine selects three bands in both blocks, extending the first block
beyond $\pm0.18$~Ha. Its measured $\eta=9.348\times10^{-14}$ gives
$s_{\mathrm{fp}}=2.979$, or $59.579$ times the decision threshold.
This large allowance is a consequence of nearly saturated occupations,
not a measured error of the candidate. Repeating the same spectra at
$\tau=0.005$~Ha gives a saturated occupation and an abstention.

Constant-potential invariance is exact algebraically but does not guarantee
floating-point invariance at arbitrary energy offsets.
The earlier implementation formed $H+\tau\operatorname{logit}(D)$ before
removing its scalar part; a large common offset could erase the
entropy-gradient differences.
The empirical allowance does not bound this loss of significance.
The corrected implementation removes a common energy origin when needed,
as tested in Sec.~\ref{sec:numerical-verification}.

Table~\ref{tab:accepted-reference-rows} lists the raw bounds, window
distances and trace deviations for each of the ten accepted
reference-configuration proposals.
For these proposals, six ratios of window error to recorded
raw bound exceed one, with a maximum of $5.414$. The largest positive
absolute excess is $1.429\times10^{-9}$; every positive excess is smaller
than its corresponding recorded trace deviation. The ratio $0.163$ belongs
to the graphene low-cutoff proposal at iteration 5 and is the maximum only
within the two graphene proposals, not the full ten-proposal set.
The raw bound alone therefore does not pass all ten numerical comparisons;
the historical verification used the empirical allowance.
This row-wise comparison does not establish the cause of the numerical
excesses or reconstruct the unsaved occupation-margin term.
It is also distinct from the particle-number correction in
Eq.~\eqref{eq:weighted-trace-defect}, which the historical implementation
did not use.
In the five accepted rows of the separate
three-system calculation,
the largest excess is $2.1\times10^{-8}$ and the corresponding trace deviation
is $5.3\times10^{-8}$. No accepted row in these records exceeds bound plus
the empirical allowance; this observation does not extend its validity to
arbitrary weights or traces.

\subsection{Stress tests and low-temperature limit}
\label{sec:stress-statistics}
The original standalone stress run retained aggregate counts only:
$3{,}586$ legal presented trials, $1{,}942$ accepted and $1{,}644$
nonaccepted. A saved, count-validated replay reproduces these totals and
the per-spectrum counts. All $3{,}586$ legal replay rows, including all
$1{,}942$ accepted rows, contain finite nonnegative reference-derived
distances. The largest accepted distance is $0.0496$ (four decimal places),
below the $0.05$ tolerance without an allowance. The largest accepted raw
distance-to-bound ratio lies $3.981\times10^{-7}$ below unity.
These replay distances use the
ordinary real-matrix Frobenius norm, not the weighted real-embedding
metric of the material-candidate tables.
The replay stores reference-derived distances, not full reference matrices
or explicit per-trial convergence flags. Matching aggregate counts does
not recover the original run's individual reference-solve status.

A false acceptance is an accepted trial
whose verified window distance exceeds the tolerance plus the empirical
allowance of Eq.~\eqref{eq:screen-slack}. No such event is reported in the
historical aggregates or the saved replay. The replay also contains 421
deliberately invalid inputs, all abstained with undefined reference distance;
they are not part of the 3,586-trial denominator. If trials
were independent, the one-sided 95\% zero-event upper bound would be
$1-0.05^{1/3586}=8.4\times10^{-4}$ per presented trial, or
$1-0.05^{1/1942}=1.5\times10^{-3}$ conditional on acceptance. Trials within
one seeded amplitude ladder share a spectrum and are correlated. Treating the
70 ladders as the independent units gives the more conservative bound
$1-0.05^{1/70}=4.2\times10^{-2}$ per ladder.

These zero-event summaries are conditional on the trial families and
numerical verification convention; finite reference distances have been
checked row by row only for the saved replay. A failed reference solve is
an unknown verification outcome,
not evidence of no violation. In the historical optional-distance path,
a missing distance could produce a flag indicating no violation.
That flag alone therefore does not establish complete verification.
The summaries are not
material-population safety probabilities.

Among replay trials whose recorded reference distance is below tolerance,
inclusion is
0.781 for the crossing spectrum and 0.828 for the metallic spectrum, or
$1{,}875/2{,}293=0.818$ pooled. The gapped spectrum gives
$67/94=0.713$; all three spectra together give
$1{,}942/2{,}387=0.814$. All $1{,}644$ rejected replay rows have a finite
recorded branch-return deviation of zero. Historical aggregate records also
report return to the initial state in 70/70 temperature-excursion tests
and 4/4 detuning round trips.

A separate seven-temperature, five-tolerance aggregate summary reports
$102{,}385$ legal trials,
with no recorded false acceptance under the same verification convention
and $5{,}655$ abstentions. It also contains
$13{,}495$ deliberately invalid inputs
with zero accepts. Double-precision occupation saturation occurs near
\begin{equation}
\tau_{\mathrm{sat}}=\frac{\mu-\varepsilon_{\min}}{36.74},\qquad
36.74=\ln(2/\varepsilon_{\mathrm{mach}}).
\end{equation}
The one-sided estimate agrees with numerically bisected boundaries to about
$10^{-4}$ relative; a two-sided estimate overpredicts them by up to a factor
2.1. Crossing spectra saturate between $1.27\times10^{-2}$ and
$1.38\times10^{-2}$~Ha, the gapped spectrum at
$8.94\times10^{-3}$~Ha, and metallic windows over
$[8.40\times10^{-3},1.13\times10^{-2}]$~Ha (median
$9.8\times10^{-3}$~Ha). The transition to systematic abstention lies about
$1.3$--$2$ times above these floors.

\section{Reference-map fallback in nonlinear SCF iterations}
\label{sec:si-loop-integration}

\subsection{Execution modes}
In the reference-scoring mode, the reference unmixed map is evaluated at every
iteration. When a proposal is accepted, the mixer nevertheless consumes the
proposal, and the simultaneously evaluated reference map is used only for
error measurement and cost accounting. In the integrated mode, the same
accepted proposal enters the mixer but the unused reference map is skipped.
All rejected or abstained proposals use the reference map. Candidate guards
reject nonfinite states, negative densities and charge drift; an empty window
produces abstention. The convergence criterion is evaluated only on iterations
that use a reference map. If the iteration limit is reached immediately after
an acceptance, one terminal reference-map evaluation is performed before a
state can be reported.

An observation-only mode augments the integrated calculation by evaluating the
skipped map after each acceptance, measuring the full-space deviation and then
discarding the result. This evaluation affects neither the decision nor the
trajectory and is excluded from every cost ratio.

Terminal agreement requires both the controlled and reference runs to
have converged and all comparison quantities to be finite.
The historical terminal comparison did not itself enforce reference
convergence or reject every nonfinite difference.
The corrected routine checks both conditions, as described in
Sec.~\ref{sec:numerical-verification}.
The finite differences reported below are observations for the historical
runs, not a general nonlinear SCF guarantee.

\subsection{Reference-configuration results}
The reference configurations use cutoffs of 3~Ha (Si and Al) and 4~Ha
(graphene), temperatures of $0.02$~Ha (Si and Al) and $0.01$~Ha
(graphene), and 8, 27 and 9 $k$ points, respectively. The low-cutoff
candidates use half the corresponding full cutoff. These configurations
are distinct from the larger 25~Ha silicon calculation below.

\begin{table}[tbp]
\centering
\caption{Reference-configuration decision counts. ``Skipped'' denotes
reference maps omitted in integrated mode. Staleness is the largest separation
between a recycled reference frame and the current iteration.}
\label{tab:reference-decisions}
\begin{tabular}{@{}lrrrrrr@{}}
\toprule
system & iterations & accept & reject & abstain & skipped & max staleness \\
\midrule
Si       & 45 & 4 & 2 & 0 & 4 & 3 \\
graphene & 69 & 2 & 2 & 2 & 2 & 3 \\
Al       & 44 & 4 & 2 & 0 & 4 & 3 \\
\bottomrule
\end{tabular}
\end{table}

No recycled-frame proposal is accepted, and no terminal verification is
needed because every run converges on a reference-map iteration. For a fixed
binary and architecture, the reference-scoring and integrated modes have
identical mixed-density trajectories and terminal states. This identity does
not imply that an accepted proposal equals the skipped map: both modes consume
the same proposal, while only one computes the unused map. The observation-only
mode reproduces the integrated trajectory and agrees with the reference-scoring
mode on all ten measured accepted-row distances.

\begin{table}[tbp]
\centering
\caption{Ten historical accepted proposals on the three reference
configurations. Candidate identifiers give the system and iteration;
L denotes low-cutoff and S denotes oracle-subspace proposals.
$B_0=r_E/(4\tau)$ is the uncorrected residual bound.
$d_W$ and $d_{\mathrm{full}}$ compare with the canonical compressed state
and the full reference-map output, respectively.
$\delta_{\mathrm{tr}}$ is the recorded absolute embedded-trace deviation,
not a universal weighted-metric correction.
Ratios are calculated before rounding. The occupation margins required
to reconstruct the complete empirical allowance were not saved.}
\label{tab:accepted-reference-rows}
\small
\setlength{\tabcolsep}{3pt}
\begin{tabular*}{\textwidth}{@{\extracolsep{\fill}}lrrrrr@{}}
\toprule
candidate & $B_0$ & $d_W$ & $\delta_{\mathrm{tr}}$ & $d_W/B_0$ & $d_{\mathrm{full}}$ \\
\midrule
Si 2 (L) & $6.405\times10^{-10}$ & $1.342\times10^{-9}$ & $2.328\times10^{-9}$ & $2.095$ & $0.540$ \\
Si 3 (S) & $1.241\times10^{-10}$ & $1.252\times10^{-10}$ & $2.292\times10^{-10}$ & $1.009$ & $0.095$ \\
Si 5 (L) & $1.138\times10^{-9}$ & $1.307\times10^{-10}$ & $2.193\times10^{-10}$ & $0.115$ & $0.475$ \\
Si 6 (S) & $2.891\times10^{-10}$ & $1.240\times10^{-10}$ & $2.207\times10^{-10}$ & $0.429$ & $0.031$ \\
graphene 5 (L) & $5.014\times10^{-11}$ & $8.193\times10^{-12}$ & $9.921\times10^{-12}$ & $0.163$ & $1.202$ \\
graphene 6 (S) & $3.624\times10^{-3}$ & $1.644\times10^{-10}$ & $1.898\times10^{-10}$ & $4.535\times10^{-8}$ & $0.403$ \\
Al 2 (L) & $3.234\times10^{-10}$ & $1.751\times10^{-9}$ & $4.083\times10^{-9}$ & $5.414$ & $0.354$ \\
Al 3 (S) & $5.245\times10^{-10}$ & $7.423\times10^{-10}$ & $1.752\times10^{-9}$ & $1.415$ & $0.007$ \\
Al 5 (L) & $3.412\times10^{-10}$ & $1.771\times10^{-9}$ & $4.140\times10^{-9}$ & $5.189$ & $0.357$ \\
Al 6 (S) & $5.207\times10^{-10}$ & $7.545\times10^{-10}$ & $1.783\times10^{-9}$ & $1.449$ & $0.005$ \\
\bottomrule
\end{tabular*}
\end{table}

The historical numerical verification used the empirical allowance in
Eq.~\eqref{eq:screen-slack}: six of these ten window distances exceed
their raw bounds, as quantified in Table~\ref{tab:accepted-reference-rows}.
Each positive excess is smaller than its corresponding recorded trace
deviation, an empirical comparison rather than an exact-trace verification.
All ten window distances remain below the operational tolerance $0.05$;
this is a different criterion from satisfying $d_W\le B_0$.
The corrected bound used in Sec.~\ref{sec:historical-materials} was not
used for these historical decisions.
Full-space
distances span $4.5\times10^{-3}$--$1.202$ (median $0.356$), whereas window
distances are at most $1.8\times10^{-9}$. Their ratio spans
$6.0\times10^6$--$1.5\times10^{11}$ on accepted rows. Thus the test does
not supply a constant relating the two metrics.

The historical small-system records give screen-to-map ratios of
0.68--4.52 and acceptance fractions of 0.333--0.667. Their integrated to
all-reference cost ratios are 1.054--1.658 (pooled 1.159), with a pooled
reference-scoring ratio of 1.226; integrated to reference-scoring ratios
are 0.928--0.978. These are legacy composite cost statistics, not ratios
of mutually exclusive wall-time intervals. In that implementation, the
screen timer started before candidate generation, while the charged total
added both screen and candidate timers. Target-orbital preparation for the oracle
candidate is also not charged as a deployment expense. The recorded
numbers are retained without a retrospective timing correction and are
not used to infer net speedup.

\subsection{Production-size silicon calculation}
\label{sec:production-silicon}
The production-size silicon calculation uses a 25~Ha cutoff, a $33^3$
real-space grid, a 64-point $k$ mesh, $\tau=0.02$~Ha, linear mixing 0.1, density and
energy tolerances $10^{-9}$ and $10^{-11}$~Ha, and proposals on iterations
2--7. Both execution modes converge in 193 iterations with density residual
$9.96\times10^{-10}$. Their terminal states and mixed-density trajectories
are identical for the fixed binary and architecture. Terminal differences are
$3.55\times10^{-15}$~Ha in free energy,
$3.00\times10^{-12}$ in normalized density and
$2.5$--$3.2\times10^{-12}$~Ha in HOMO/LUMO energies, all below the stated
comparison tolerances. A separately built reference calculation converges one
iteration later because the same residual threshold is crossed at
$9.13\times10^{-10}$ rather than $9.96\times10^{-10}$; the terminal free
energies agree at printed precision.

Four of six proposals are accepted, and two coarse-mesh proposals are rejected.
The rejected proposals have bounds $6.79$--$6.97$ and directly evaluated
window distances
$0.372$--$0.379$, compared with tolerance 0.05. Accepted bounds span
$7.4\times10^{-10}$--$3.3\times10^{-9}$ and measured window distances
$4.9\times10^{-11}$--$4.2\times10^{-10}$.

\begin{table}[tbp]
\centering
\caption{Accepted rows in the production-size silicon calculation.}
\label{tab:production-accepted}
\begin{tabular}{@{}rrll@{}}
\toprule
iteration & proposal & window distance & full distance \\
\midrule
2 & low-cutoff & $3.22\times10^{-10}$ & $0.022$ \\
3 & oracle subspace & $4.16\times10^{-10}$ & $0.240$ \\
5 & low-cutoff & $1.92\times10^{-10}$ & $0.021$ \\
6 & oracle subspace & $4.93\times10^{-11}$ & $0.161$ \\
\bottomrule
\end{tabular}
\end{table}

Two accepted full-space distances exceed 0.05 even though all window distances
are of order $10^{-10}$. This is the same local-to-global separation observed
in the smaller calculations.

\begin{table}[tbp]
\centering
\caption{Historical timing records for the larger silicon calculation.
The in-loop totals are legacy composite costs with potentially overlapping
timers, not mutually exclusive wall-time measurements. The isolated screen
measurement is from a separate calculation, not the integrated run.}
\label{tab:production-cost}
\begin{tabular}{@{}lr@{}}
\toprule
quantity & value \\
\midrule
reference map, in loop & $11.1$ s per call \\
screen, serial in-loop implementation & $261$ s per proposal \\
proposal generation, serial in-loop implementation & $2.21\times10^3$ s per proposal \\
integrated/all-reference legacy composite ratio & $7.87$ \\
reference-scoring/all-reference legacy composite ratio & $7.93$ \\
integrated/plain-reference legacy composite ratio & $7.05$ \\
reference-scoring/plain-reference legacy composite ratio & $7.10$ \\
in-loop screen/map ratio & $23.5$ \\
separately measured production-path screen/map ratio & $0.170$ \\
\bottomrule
\end{tabular}
\end{table}

The two legacy composite totals differ by $136$~s. This difference does
not measure wall time saved by skipping maps, and the historical timers
cannot be made mutually exclusive without establishing their exact source
version and measurement boundaries. The recorded in-loop screen-to-map
ratio is $23.5$, with a per-proposal range $19.4$--$31.3$.
The separately measured production-path ratio is
$r=1.699/10.013=0.170$ to the displayed precision, obtained by timing the
production screen and reference map in the same separate calculation.
This isolated ratio does not include proposal generation or other
per-proposal work, and it is not an end-to-end speed measurement.
It therefore cannot replace the in-loop ratio in the historical cost totals.
The legacy totals show no cost reduction under their accounting convention;
they establish neither a corrected slowdown factor nor a net speedup.
Only six proposals were scheduled, on iterations 2--7, and four were accepted.
For this schedule, an unchanged trajectory with equal-cost maps and free
candidates and screening would have speedup $193/189\simeq1.021$.
Even accepting all six scheduled proposals would give
$193/187\simeq1.032$ under the same assumptions.
These estimates describe this schedule, not a general limit of the method;
they do not apply when map costs or the trajectory change.

\subsection{Enlarged in-loop sample}
\label{sec:enlarged-sample}
The enlarged calculation contains 48 additional runs on Si, graphene and Al,
with one proposal inserted per run. Low-cutoff proposals are inserted at
iterations $\{2,5,8,11,14,17\}$, oracle-subspace proposals at
$\{3,6,9,12,15,18\}$ and coarse-$k$ proposals at
$\{4,7,10,13\}$. The 48 runs yield 30 accepts, 12 rejects and 6 empty-window
abstentions. All coarse-$k$ proposals are rejected. Combining these runs with
the initial reference-configuration sample gives 40 accepted, 18 rejected and
8 abstained proposals; 58 rows have defined window distances and 35 accepted
blocks have a subsequent reference-map iteration.

The bound-based decision agrees with the directly evaluated window-distance
decision on all
58 rows. The smallest rejected and largest accepted window distances are
separated by a factor $1.8\times10^8$. In contrast, full-space distances give
a pooled AUC of $0.694$.
Here AUC denotes the fraction of accepted--rejected pairs in which the
rejected candidate has the larger full-space distance, with half weight
assigned to ties.
There are no ties across the two decision groups in these records.
Table~\ref{tab:auc-pairs} resolves the $500$ concordant pairs among
$40\times18=720$ pairs by material.
The same-material comparisons give $156/240=0.650$, whereas the
cross-material comparisons give $344/480\simeq0.717$.
The pooled value is their pair-count-weighted mean,
\begin{equation}
 \operatorname{AUC}_{\mathrm{pool}}
 =\frac{500}{720}
 =\frac13\cdot\frac{156}{240}+\frac23\cdot\frac{344}{480}
 \simeq0.694.
 \label{eq:auc-decomposition}
\end{equation}

\begin{table}[tbp]
\centering
\caption{Full-space distance ordering by material in the enlarged sample.
Each entry is the number of concordant accepted--rejected pairs divided
by the total number of pairs.
Rows identify the accepted candidate's material and columns identify the
rejected candidate's material. Diagonal entries are same-material
comparisons; off-diagonal entries are cross-material comparisons.}
\label{tab:auc-pairs}
\begin{tabular*}{0.85\textwidth}{@{\extracolsep{\fill}}lrrr@{}}
\toprule
& \multicolumn{3}{c}{Rejected material} \\
\cmidrule(l){2-4}
Accepted material & Si & graphene & Al \\
\midrule
Si & $48/96$ & $80/96$ & $96/96$ \\
graphene & $0/48$ & $12/48$ & $24/48$ \\
Al & $48/96$ & $96/96$ & $96/96$ \\
\bottomrule
\end{tabular*}
\end{table}

Per-system AUCs are $0.500$ (Si), $0.250$ (graphene) and $1.000$ (Al).
Thus the full-space ordering varies by material; it does not disappear
when comparisons are restricted to the same material.
All 18 rejected candidates are coarse-$k$ proposals, whereas all 40
accepted candidates are low-cutoff or oracle-subspace proposals.
Decision and proposal family are therefore confounded in this sample.
The AUC does not isolate the screen's effect or test whether the
continuous residual bound ranks full-space errors.

Independently resampling accepted and rejected rows $10{,}000$ times gives
percentile endpoints $[0.544,0.831]$
(2.5th and 97.5th percentiles; seed 20260829).
This row-level resampling ignores dependence within materials and proposal
sequences; it does not estimate uncertainty across independent materials.
Deleting Si, graphene or Al gives pooled AUCs $0.792$, $0.750$ and $0.486$,
respectively.
These are deletion-sensitivity calculations, not prediction tests on
held-out materials.
Accepted and rejected full distances overlap on $[0.325,0.952]$.
The largest accepted value, $1.202$, exceeds the largest rejected value,
$0.952$, despite the perfect separation within the Al sample.

The contraction analysis uses the Euclidean mixed-density increment
$u_n=\|\rho_n^{\mathrm{mix}}-\rho_{n-1}^{\mathrm{mix}}\|_2$ on the fixed
real-space grid. For each consecutive accepted block, $j$ is the first later
iteration that evaluates a reference map.
We fit $\log_{10}u_n=a+bn$ over the 12 iterations following the last scheduled
proposal, retaining positive increments with reference maps at both adjacent
iterations. All measured post-block iterations $j$ are excluded from the fit.
The ratio $P=u_j/10^{a+bj}$ therefore uses subsequent trajectory data and
describes contraction retrospectively; it is not an online forecast.
It also differs from the unmixed Kohn--Sham fixed-point residual.
For the 35 accepted blocks, $P$ spans $[0.594,1.312]$ (median 1.046).
Spearman correlation between $\log P$ and the log window distance is $-0.349$
with an approximate Fisher-transformed interval $[-0.611,-0.018]$;
per-system values are $+0.495$
(Si), $-0.107$ (graphene) and $+0.521$ (Al). The pooled sign is therefore not
a common system-level relation. Correlation with full distance is $+0.708$
[$+0.490,+0.842$], but both quantities depend strongly on insertion position
within the subspace sequence: position correlations are $-0.921$ for full
distance and $-0.871$ for $P$. The apparent positive relation therefore
contains a contraction-phase dependence. Correlation between $\log P$ and the
log bound is $+0.221$ with interval $[-0.121,+0.516]$.
These correlation intervals use
$\tanh[\operatorname{atanh}(\widehat\rho)\pm1.960/\sqrt{n-3}]$ with $n=35$;
they do not account for shared-system dependence.
The per-system results and insertion-position comparisons establish the
descriptive scope of these associations. These results support
the distinction between the window-level bound and subsequent nonlinear SCF
dynamics.

\section{Historical-subspace material comparisons}
\label{sec:historical-materials}

Three additional calculations tested candidates without the complete solution
of the current Hamiltonian. The structures and numerical parameters are given
in Table~\ref{tab:historical-settings}. These small discretizations test the
screening procedure; they are not new cutoff-convergence calculations.
All calculations used analytic HGH pseudopotentials, LDA-PZ81, and
spin-degenerate Fermi--Dirac occupations. The solver used complex dense
Hamiltonians, faer diagonalization, and direct plane-wave density reconstruction.

\begin{table}[hbp]
\centering
\caption{Settings for the historical-subspace comparisons. Cell vectors and
pseudopotential parameters are included in the numerical reconstruction data.
Meshes use Monkhorst--Pack sampling without an additional offset.}
\label{tab:historical-settings}
\begin{tabular}{lcccccc}
\toprule
system & atoms & $N_e$ & cutoff (Ha) & k mesh & FFT grid & $(\tau\,[\mathrm{Ha}],\alpha)$ \\
\midrule
diamond Si & 2 & 8 & $3.000$ & $2^3$ & $11^3$ & $(0.020,0.300)$ \\
graphene & 2 & 8 & $4.000$ & $3\times3\times1$ & $9\times9\times37$ & $(0.010,0.200)$ \\
fcc Al & 4 & 12 & $3.000$ & $3^3$ & $11^3$ & $(0.020,0.300)$ \\
\bottomrule
\end{tabular}
\end{table}

The initial density was $\rho_1=N_e/\Omega$.
One reference map produced its output density and complete orbitals $U_1$.
Linear mixing gave $\rho_2=(1-\alpha)\rho_1+\alpha\rho_1^{\rm out}$,
followed by roundoff-level charge normalization.
At each k point, the retained subspace ended exactly two bands above the
last $U_1$ occupation greater than $10^{-12}$.
Rayleigh--Ritz diagonalization of $H[\rho_2]$ in that subspace and a shared
chemical potential supplied the candidate occupations.
The candidate density matrix was zero on the omitted orthogonal complement.
No orbitals from a converged target calculation were used.

The candidate, current Hamiltonian, and screening decision were saved before
the complete current-Hamiltonian reference was evaluated.
Window selection counted levels within $\pm18\tau$ at each k point and used
the maximum count for all blocks. An empty block or insufficient retained
bands caused abstention, with no change to the specified window rule.
The window target subtracted actual excluded Fermi occupations, including
their fractional tails. This differs from the integer frozen-charge
convention in the earlier controller tests.
The corrected bound was
\begin{equation}
B_W=\frac{r_E}{4\tau}+\frac{|\delta N_e|}{\sqrt{2w_{\min}}},
\label{eq:historical-material-corrected-bound}
\end{equation}
where $r_E$ is the residual in the real-embedded weighted metric.
The candidate charge gate remained $10^{-6}$ electrons and the window
distance threshold remained $0.05$.
All post-decision reference charge errors were checked against
$10^{-10}$ electrons.

Si and Al were accepted, with corrected bounds and measured window distances
reported in the main text. Graphene abstained because some k points had
no levels in the prescribed interval. Its candidate remained well defined
and was scored against the complete reference.
There were no failed or missing material points and no repeated attempts.
The full fixed-$H$ free-energy differences and the resulting next-input
differences are listed in Table~\ref{tab:historical-diagnostics}.

\begin{table}[hbp]
\centering
\caption{Complete-state diagnostics for the three historical-subspace candidates.
$\Delta F_H$ is the signed spin-degenerate fixed-Hamiltonian free-energy
difference, not a Kohn--Sham total-energy difference.
$\epsilon_{\rho,\mathrm{next}}$ compares the two raw linear mixed inputs.
The last column measures the scalar identity
$\epsilon_{\rho,\mathrm{next}}=\alpha\epsilon_\rho$; it is not a nonlinear
convergence test.}
\label{tab:historical-diagnostics}
\begin{tabular}{lccc}
\toprule
system & $\Delta F_H$ (Ha) & $\epsilon_{\rho,\mathrm{next}}$ & identity difference \\
\midrule
diamond Si & $5.652\times10^{-3}$ & $0.011$ & $-1.735\times10^{-18}$ \\
graphene & $1.267\times10^{-2}$ & $0.012$ & $5.204\times10^{-18}$ \\
fcc Al & $8.596\times10^{-6}$ & $3.605\times10^{-4}$ & $-5.421\times10^{-19}$ \\
\bottomrule
\end{tabular}
\end{table}

The energy comparison inside a window used a separate canonical reference
at the candidate's measured trace. The difference from a reference at the
nominal target trace was retained only as a diagnostic.
The distance and same-trace energy comparisons used the fixed numerical
allowance $10^{-10}\max(1,|\mathrm{lhs}|,|\mathrm{rhs}|)$.
This allowance tests floating-point consistency and does not enclose
rounding error rigorously.

An independent NumPy implementation repeated the canonical diagonalizations
and plane-wave density reconstruction from the saved Hamiltonians and
candidates. All 1,135 numerical consistency checks passed, using absolute
and relative comparison tolerances of $10^{-10}$ and $10^{-9}$, respectively.
The largest embedded distance between the two complete reference density
matrices was $2.742\times10^{-13}$.
The largest change in the reported full candidate distance was
$5.69\times10^{-15}$. These checks confirm the fixed-$H$ comparisons;
they do not alter the previously recorded screening decisions.
The nearly stationary window distances depend on reference-solve accuracy:
the independent references gave $3.656\times10^{-15}$ for Si and
$1.426\times10^{-14}$ for Al. These values and the stored-reference
distances in the main table all lie in the near-zero numerical regime.

Each material used two reference maps, for six maps in total.
The first map obtains spectra and orbitals separately, so a map count is
not an eigensolver-call count. The additional Hamiltonian consistency checks,
Ritz solves, small-window references, density reconstruction, and saved-record
operations were included in the diagnostic timings.
The three single-threaded runs took $10.914$, $19.123$, and $81.351$~s,
respectively, on an x86-64 Linux compute node using Rust 1.96.0.
These times describe the complete comparison procedure, not time saved in SCF.
No subsequent nonlinear map was evaluated, and no conclusion about terminal
SCF accuracy or acceleration is inferred from these three points.

\section{Classical mixing comparison}
\label{sec:mixing-comparison}

The operating points are as follows. Diamond Si: FCC primitive cell,
HGH-Si LDA-PZ81, $\Gamma$ sampling, 3.5~Ha cutoff, $\tau=0.02$~Ha, linear
parameter 0.1, density/energy tolerances $10^{-7}/10^{-9}$ and 400 iterations.
Graphene: 20-bohr vacuum, shifted $6\times6\times1$ mesh containing K, 10~Ha
cutoff, $\tau=0.01$~Ha, linear parameter 0.2, tolerances
$10^{-6}/10^{-8}$ and 150 iterations. FCC Al: $2\times2\times2$ mesh,
2.5~Ha cutoff, $\tau=0.02$~Ha, linear parameter 0.3, tolerances
$10^{-7}/10^{-9}$ and 250 iterations.

The five policies are linear mixing; Kerker with $q_0=1.0$~bohr$^{-1}$;
Pulay with history 6, regularization $10^{-10}$, $|c_i|\leq4.0$ and a
1.5-linear-step limit; the same Pulay scheme on the Kerker map; and Pulay
without this limit.

The Pulay regularizer is a fixed absolute value. As the residual Gram
matrix decreases in scale, this term can dominate and drive history
coefficients toward averaging. These are comparisons of the specified
parameter choices, not of optimally tuned Pulay, DIIS or Kerker methods.

\begin{table}[tbp]
\centering
\caption{Classical mixing results. Wall times are min--max ranges over three
repetitions (s); ``rejected'' indicates violation of density non-negativity in
every repetition.}
\label{tab:mixing-comparison}
\begin{tabular}{@{}llrrr@{}}
\toprule
system & policy & outcome/iterations & wall (s) & clamps \\
\midrule
Si & linear & 152 & $0.80$--$0.83$ & -- \\
 & Kerker & 247 & $1.31$--$1.35$ & -- \\
 & Pulay, clamped & 351 & $1.89$--$1.91$ & 67 \\
 & Pulay on Kerker & limit at 400 & $2.14$--$2.16$ & 106 \\
 & Pulay, unclamped step & 264 & $1.41$--$1.43$ & 35 \\
\midrule
graphene & linear & 69 & $143.9$--$263.7$ & -- \\
 & Kerker & rejected & -- & -- \\
 & Pulay, clamped & 134 & $338.4$--$374.8$ & 40 \\
 & Pulay on Kerker & rejected & -- & -- \\
 & Pulay, unclamped step & rejected & -- & -- \\
\midrule
Al & linear & 41 & $0.90$--$0.95$ & -- \\
 & Kerker & 58 & $1.27$--$1.29$ & -- \\
 & Pulay, clamped & 94 & $1.97$--$2.07$ & 19 \\
 & Pulay on Kerker & 136 & $2.78$--$2.81$ & 24 \\
 & Pulay, unclamped step & 67 & $1.44$--$1.47$ & 0 \\
\bottomrule
\end{tabular}
\end{table}

Initial state and stopping criteria are identical within each system.
All 11 converged cases reach the linear-mixing fixed point. The maximum
$(|\Delta F|,d_\rho)$ values are
$(5.3\times10^{-15}\ \mathrm{Ha},1.2\times10^{-9})$ (Si),
$(1.4\times10^{-13}\ \mathrm{Ha},1.1\times10^{-11})$ (graphene) and
$(2.0\times10^{-14}\ \mathrm{Ha},1.4\times10^{-10})$ (Al). Si Pulay on
Kerker reaches the iteration limit at
$(|\Delta F|,d_\rho)=(7.0\times10^{-13}\ \mathrm{Ha},3.8\times10^{-8})$;
only the energy criterion is stationary.

For the rejected graphene verification runs, the last recorded iterations
are 3 (Kerker), 3 (Pulay on Kerker) and 5 (unclamped Pulay); these are completed
iteration records, not the indices of the subsequent rejected density maps.
Unclamped Pulay reduces the residual from 0.41 to 0.33 before rejection;
clamped Pulay converges in 134 iterations with 40 clamps. Removing
the Pulay step limit reduces iterations by factors 1.33 (Si) and 1.40 (Al), but
requires 1.74 and 1.63 times as many iterations as linear mixing. The reported
successful iteration counts and failure outcomes repeat across three timing
runs; rejection iteration indices were not recorded for those repetitions.
Timings compare only within a system; graphene shows substantial load variation.

\bibliography{references}